# Pilot Quantum Error Correction for Global-Scale Quantum Communications

Laszlo Gyongyosi[*,1,2], *Member, IEEE,* Sandor Imre[1], *Member, IEEE*

[1]*Quantum Technologies Laboratory, Department of Telecommunications*

*Budapest University of Technology and Economics*

2 Magyar tudosok krt, H-1111, Budapest, Hungary

[2]*Information Systems Research Group, Mathematics and Natural Sciences*

*Hungarian Academy of Sciences*

H-1518, Budapest, Hungary

[*]*gyongyosi@hit.bme.hu*

Real global-scale quantum communications and quantum key distribution systems cannot be implemented by the current fiber and free-space links. These links have high attenuation, low polarization-preserving capability or extreme sensitivity to the environment. A potential solution to the problem is the space-earth quantum channels. These channels have no absorption since the signal states are propagated in empty space, however a small fraction of these channels is in the atmosphere, which causes slight depolarizing effect. Furthermore, the relative motion of the ground station and the satellite causes a rotation in the polarization of the quantum states. In the current approaches to compensate for these types of polarization errors, high computational costs and extra physical apparatuses are required. Here we introduce a novel approach which breaks with the traditional views of currently developed quantum-error correction schemes. The proposed solution can be applied to fix the polarization errors which are critical in space-earth quantum communication systems. The channel coding scheme provides capacity-achieving communication over slightly depolarizing space-earth channels.

# I. Introduction

Quantum error-correction schemes use different techniques to correct the various possible errors which occur in a quantum channel. In the first decade of the 21st century, many revolutionary properties of quantum channels were discovered [12-16], [19-22] however the efficient error-correction in quantum systems is still a challenge. These error-correction schemes use different techniques to correct the various possible errors which can occur in a quantum channel. One of the biggest challenges in current quantum error-correction techniques is redundancy [8]. To protect an arbitrary quantum state it has to be encoded in a redundant way, and redundant encoding was an unavoidable corollary in quantum error-correction. The level of redundancy can differ, however one issue is that it cannot be removed from currently known approaches [1-5], [7-9]. A second major problem is that the various types of errors which can occur in the quantum channel require different encoding and decoding schemes, moreover if the type of the error is completely unpredictable then the only way to achieve protection of the quantum state is to increase the level of redundancy [9-12].

In this paper we introduce a new quantum-error correction scheme which has many advantages in comparison to currently known techniques. First of all, the proposed method requires minimal redundancy. Second, the error correction can be made without any knowledge about the transformation of the quantum channel. The proposed error correction scheme is based on the usage of *pilot* states. The pilot states are ordinary quantum states, fed by Alice to the quantum channel. The pilot states will capture and store the unknown error transformation of the quantum channel. These pilot states—using a simple Hadamard and Controlled-NOT (CNOT) gates [4]—can be used to correct an arbitrarily high number of data quantum states sent through the quantum channel. However the errors on all of these states are unknown, Bob is able to construct them using our simple quantum circuit. Furthermore, the quantum channel is stored in the pilot quantum state without making process tomography on the channel. The simplicity of the proposed error-correction quantum circuits makes it possible to be easily implemented in practice. The hard part is to construct an error-correction scheme which can correct the unknown errors of the channel, without any knowledge of the channel output pilot states. The probabilistic behavior of these controlled quantum gates, and the storing process of unitary transformations in quantum states were studied in the literature [12,13].

An immediate practical application of the pilot quantum-error correction is in *polarization compensation in space-earth quantum communications* [1], [6], [14-16]. In space-earth quantum communications the relative motion of the ground station and the satellite causes a rotation in the polarization of the quantum states. In the current approaches to compensate for these types of polarization errors, high computational costs and extra physical apparatuses are required. The proposed quantum error-correction scheme can be applied to fix the polarization errors which are critical in space-earth quantum communication systems. Our polarization compensation scheme can be implemented in practice without any extra hardware or software costs, providing an easily implemented on-the-fly polarization compensation scheme.

There are several polarization techniques existing in the literature, but in each case, further hardware and software implementations are required and the cost of these practical polarization solutions is high. Our scheme uses minimal redundancy and adds it into an arbitrary large block set instead of each individual qubits. Using our scheme, the error-correction of the data qubits can be achieved without any knowledge about angle of the polarization rotation error.

This paper is organized as follows. In Section 2, we give an exact problem in space-earth quantum channels where the proposed technique can be applied. In Section 3 we show the details of the channel coding scheme. In Section 4 we introduce the channel model and express the capacities. In Section 5 we present the theorems and the proofs. In Section 6 a practical implementation with capacity calculations for space-earth quantum channels is shown. Finally, in Section 7 we conclude the results.

## II. Polarization Compensation

### A. Problem Statement

In space-earth quantum communications the relative motion of the ground station and the satellite causes a rotation in the polarization of the quantum states as depicted in Fig. 1. The relative motion of the ground station and the satellite is denoted by $\theta$. During time $T$ the ground station sends $n$ data qubits and the relative motion causes a unitary rotation $U_\theta$ according to the angle $\theta$. The time-dependent rotation angle $\theta(T)$ is constant for a given time $T$ (for realistic results see Table 2.). The rotation causes time-dependent $U_\theta(T)$ unitary transformation of the polarization states of the sent qubits.

In time-domain $T \in \left[0, T_{crit.}\right]$ the channel is modeled by the same map, where $T_{crit.}$ is the critical upper bound on time parameter $T$ during the quantum channel is in the "*stationary state*". The pilot error-correction method corrects for the rotation caused by the rotation angle $\theta(T)$, without any redundant qubits, extremal hardware devices or any complex mathematical calculations.

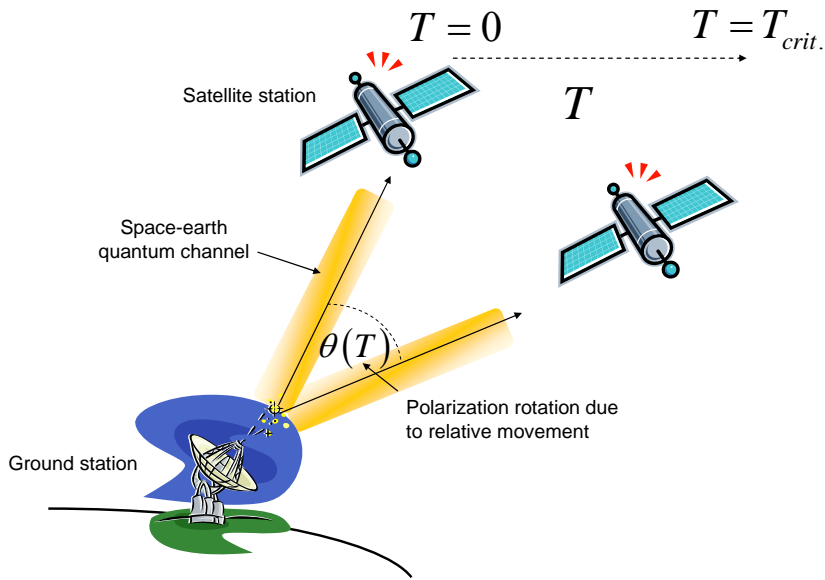

**Figure 1.** During time $T$, the relative movement causes a unitary rotation $\theta(T)$ in the polarization of the qubits.

In existing *polarization control schemes* [15], the relative motion is compensated in several steps, requiring many extra hardware and software costs. In current approaches, the most important part of these correction steps is the calculation of the time-dependent Jones matrix $C(T)$, which describes the rotation of the polarization angles in the form of Jones vectors

$$J = \begin{bmatrix} a \\ be^{i\varphi} \end{bmatrix}, \quad (1)$$

where $a^2 + b^2 = 1$ are the relative amplitudes, $\varphi$ is the relative phase ($a, b, \varphi$ are real values) and

$$J(T) = C(T)J_0, \quad (2)$$

where $T$ is the time-slot during the rotation with the given angle $\theta$ occurring on the polarization of the qubits, while $J_0$ is the initial polarization state sent by Alice in the beginning of the communication [15].

## B. Related Work

The aim of the existing solutions is the description of the Jones matrix $C(T)$, which requires the measure of the space-earth quantum channel. It is a very error-sensitive process, since during the measuring process the angles of the original qubits cannot be perturbed, i.e., it has to be implementable in practice without any physical contact between the polarization angles and the measuring beam [15]. In current approaches, the measurement process requires many extra costs, such as the use of ultra-sensitive hardware devices which can characterize the properties of the quantum channel with a *probe-beam* using a special wavelength (different from the signal) without causing any interference. Aside from experimental challenges such as the measure of the quantum channel, the mathematical derivation and calculation of the time-dependent Jones matrix causes further increase of the costs of the error-correcting scheme [15].

Another practical approach for polarization compensation is the *time-multiplexing of signal and probe beam*, which technique uses the same wavelength for the signal beam and the probe beam. In either case, the implementation costs are high and cannot be decreased, since these are arising from the characteristics of the physical apparatuses used for the polarization compensation [15]. The pilot quantum error-correction scheme does not require the characterization of the Jones matrix $C(T)$ and does not require any computations regarding it, nor does it require the implementation of any extra devices for the probe beaming, the calibration of the wavelength, or the delicate channel measurement strategy. The developed technique provides an *on-the-fly* solution for correcting polarization errors, without any matrix computations or the usage of any probe beaming devices.

## III. Coding Scheme

The proposed method can correct any unitary error $U$ in the polarization angles of the data qubits without using any channel-measuring process or probing beams, or causing any disturbance in the satellite quantum channel. The extreme efficiency of the on-the-fly strategy is ensured by two main attributes of a satellite communication system.

The number of maximally transmittable and correctable qubits depends on the length of the time $T$ during the rotation $\theta$ occurring on the polarization of the qubits in the space-earth channel, and on the frequency at which the incoming beam is modulated [15, 16]. Current practical approaches all make it possible to send as many qubits as required for the proposed polarization compensation [15], see Table 1 and Table 2. (*Note*: The results will be demonstrated for qubit inputs ($d$=2 dimensional systems) and qubit channels.)

### A. Error Characterization

The unknown *polarization rotation* (constant for time $T$, see Fig. 1 and Table 2 [16]) due to the relative movement is expressed by

$$U_\theta = e^{i\frac{\theta}{2}G}, \tag{3}$$

where $G$ is a *Hermitian generator operator* (i.e., $G = G^{\dagger}$, its eigenvalues are real and it is also a normal operator) of arbitrary dimension and $\theta$ is a real number. (*Note*: The map of the realistic space-earth channel will be given in Eq. (14)). Any Hermitian operator $G$ can be expressed as a linear combination of the Pauli operators $I$, $X$, $Y$ and $Z$. As follows, the pilot quantum error-correction scheme can correct $I$ (Identity transformation, i.e., no error), plus any single-qubit $X$, $Y$, or $Z$ error [11, 12], therefore it can correct an *arbitrary* single-qubit error, including *non-unitary* ones. It can be verified easily, because $I$, $X$, $Y$, and $Z$ form a basis for the space of 2x2 matrices, and every single-qubit error can be described by a 2x2 matrix [1-5], [15,16]. If the generator matrix $G$ is also *unitary*, i.e., $G^{-1} = G^{\dagger}$ then

$$G = G^{\dagger} = G^{-1} \rightarrow GG^{\dagger} = GG = I, \quad (4)$$

thus $G$ is self-inverse, thus $G^2 = I$. As follows, the error transformation with the self-inverse generator $G$ (the Pauli operators are also self-inverse operators) of the quantum channel can be rewritten as

$$U_{\theta} = \cos\frac{\theta}{2}I + i\sin\frac{\theta}{2}G. \quad (5)$$

To reveal and store this channel operation in a quantum state, we will use pilot quantum states to describe the transformation of the channel. The continuous variable $\theta$ will be stored in the quantum state. We note, *non-unitary* errors also can be corrected with this technique (using multiple pilot states for the decoding), since any Hermitian generator matrix $G$ can also can be stored in this way [1-4]. In the further parts of the paper we consider the case if the generator matrix $G$ is also unitary (self-inverse), i.e., its inverse can be implemented in practice by a controlled unitary transformation $U^{\dagger}$. On the other hand, our results make possible to store and correct *any non-unitary errors* [16]. In the proposed pilot error-correcting scheme the unitary operator $U_{\theta}$ (see Eq. (5)) is stored in a quantum state, which problem is equal to store a continuous variable $\theta$ in a single qubit [12,13]. *Note*: the term *unknown* data state will refer to the encoding of *quantum information*, while for the transmission of *classical information* the term *known* data state will be used. We also use the term *unknown* for the pilot states, which refers to the unknown probability amplitudes.

The input pilot state $\left|\varphi\right>_{IN}^{Pilot}$ is a quantum state, sent by Alice to the quantum channel. The $U_\theta$ unknown transformation of the channel will transform the input pilot state into

$$\left|\varphi\right>_{OUT}^{Pilot} = U_\theta \left|\varphi\right>_{IN}^{Pilot} = \left|\theta\right>. \tag{6}$$

The quantum channel $\mathcal{N}$ during time $T$ carries out the same rotation transformation (according to the relative movement, see Fig. 1) on all of the pilot and data qubits [16]. The map of the quantum channel is characterized by the self-inverse operator $G$, will be produced on the input pilot state.

## B. System Characterization

The input pilot state cannot be an eigenvector of the Pauli $X$, $Y$ and $Z$ operators, which excludes the following states:

$$\begin{aligned} X &= \left\{ \frac{1}{\sqrt{2}} (\left|0\right> \pm \left|1\right>) \right\}, \\ Y &= \left\{ \frac{1}{\sqrt{2}} (\left|0\right> \pm i\left|1\right>) \right\}, \\ Z &= \{\left|0\right>, \left|1\right>\}. \end{aligned} \tag{7}$$

Furthermore, the pilot state cannot be a pure state, since every quantum state is an eigenstate of some unitary operator, henceforth a fixed pure quantum state cannot work for all errors. As follows, the *pilot state can be any mixed, but non-maximally mixed state*, i.e., the only condition on the mixed input pilot state $\rho_{IN}^{Pilot}$ is the following:

$$\rho_{IN}^{Pilot} \neq \frac{1}{2} I. \tag{8}$$

This trivially follows from that fact, that $\rho_{IN}^{Pilot}$ has to store only the rotation caused by the channel, and an arbitrary, non maximally-mixed state can be chosen for this purpose.

The result will be the pilot state

$$\mathcal{N}\left(\rho_{IN}^{Pilot}\right) = U_{\theta}\rho_{IN}^{Pilot}U_{\theta}^{\dagger} = \sigma_{OUT}^{Pilot}. \tag{9}$$

The state $\sigma_{OUT}^{Pilot}$ is a mixed state, however, in the calculations we can use the $|\theta\rangle$ purification state of this system, which defines a pure system. Bob will apply the Hadamard transformation on the pure system $|\theta\rangle$, which is given by

$$\sigma_{OUT}^{Pilot} = \mathcal{N}\left(\rho_{IN}^{Pilot}\right) = Tr_E |\theta\rangle\langle\theta|, \tag{10}$$

where

$$|\theta\rangle\langle\theta| = \left(\cos\frac{\theta}{2}|0\rangle + i\sin\frac{\theta}{2}|1\rangle\right)\left(\cos\frac{\theta}{2}\langle 0| + i\sin\frac{\theta}{2}\langle 1|\right), \tag{11}$$

and $E$ is the environment. As follows, the pure $|\theta\rangle$ channel output pilot state in (11) is the *purification state* of the mixed channel output state $\sigma_{OUT}^{Pilot}$, and we can calculate with system $|\theta\rangle$ in the further steps.

**Comparing of Output Pilot State with the Reference Pilot State**

Before the communication, Alice and Bob agree on a pure *reference* pilot state $|\varphi\rangle_{\text{REF.}}^{Pilot}$, which can state is an arbitrary pure state. As follows, for an $I$ identity channel, Bob will not rotate the channel output states, since in that case Bob will find that for the purified state $|\theta\rangle$ the following relation holds

$$|\theta\rangle = |\varphi\rangle_{\text{REF.}}^{Pilot}, \tag{12}$$

which easily can be verified by a SWAP-test circuit [37] (see Theorem 1 in Section 5). Applying it on Bob's side, he will be able to determine easily whether he has to apply the rotation or not, by simply comparing the purification of $\mathcal{N}\left(\rho_{IN}^{Pilot}\right)$, denoted by $|\theta\rangle$, with his pure reference pilot state $|\varphi\rangle_{\text{REF.}}^{Pilot}$.

Upon reception of the pilot states, Bob can use them to correct the errors of the quantum channel, see the theorems and proofs of Section 5. The process of the pilot state generation is summarized in Fig. 2. The input pilot qubit is a well-characterized and exactly chosen qubit (and known a-priori by the parties, referred as $|\varphi\rangle_{\text{REF.}}^{Pilot}$). The channel output pilot state captures the unknown map of the channel. (The $G$ generator matrix of the channel will be defined in Section 5, see (52).)

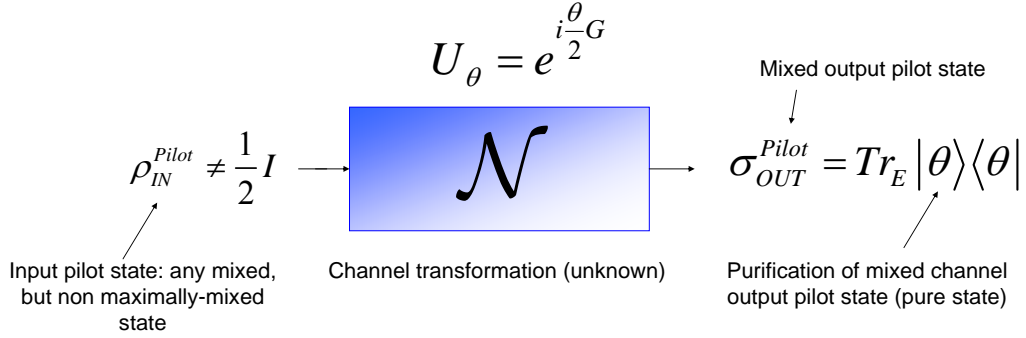

**Figure 2**. In the transmission process, Alice sends the pilot qubits to Bob before the data qubits. These states will store the unknown transformation of the quantum channel.

After the purification of $\sigma_{OUT}^{Pilot}$, the $U_\theta$ unknown unitary polarization rotation transformation of $\mathcal{N}$ can be described in the following *channel output pilot quantum* state

$$|\theta\rangle = \cos\frac{\theta}{2}|0\rangle + i\sin\frac{\theta}{2}|1\rangle, \tag{13}$$

which means that the unknown unitary transformation is mapped onto the state $|1\rangle$, while - provided the channel has no error and hence realizes an identity transformation $I$ - is mapped onto $|0\rangle$ [12,13]. After the unknown transformation of the quantum channel is stored in the pilot quantum state $|\theta\rangle$, we would like to use it in quantum-error correction, but without adding any redundancy into the encoding process [16].

## IV. The Time-Dependent Depolarizing Quantum Channel

In the previous section the quantum channel is assumed to be an arbitrary channel map, which does a time-dependent unitary rotation transformation $U_\theta$. This error occurred by the relative movement and caused an angle $\theta$ theta rotation in the polarization states. The space-earth quantum channels have no absorption since the signal states are propagated in empty space, on the other hand a small fraction of these channels is in the atmosphere, which causes slight depolarizing effect [1], [6], [14-16], [35-36]. To make our channel model more realistic, we can assume that this channel is a *time-dependent depolarizing channel*, which will be denoted by $\mathcal{N}_{depol.}^T$. This channel behaves as a standard depolarizing channel [30-33] for $T \in [0, T_{crit.}]$, (and which channel applies also the $U_\theta$ rotation operator) while it becomes a completely depolarizing channel if $T > T_{crit.}$, where $T_{crit.}$ is the critical upper bound on time parameter $T$ during channel $\mathcal{N}_{depol.}^T$ is in the "*stationary state*" (see Fig. 1). The time-dependent $\mathcal{N}_{depol.}^T$ channel has positive capacity only in time domain $T \in [0, T_{crit.}]$. If $T > T_{crit.}$, the capacities become zero, i.e., $C(\mathcal{N}_{depol.}^T) = Q(\mathcal{N}_{depol.}^T) = 0$. (*Note*: The depolarizing channel models a worst-case scenario assuming an extremely noisy atmospheric environment. The actual choice of the corresponding channel model could depend on the weather conditions, environment properties and other atmospheric parameters. These environment-specific attributes determine the exact channel model.)

For $T \in [0, T_{crit.}]$, channel $\mathcal{N}_{depol.}^T$ is a unital channel, and the CPTP (*Completely Positive Trace Preserving*) map of $\mathcal{N}_{depol.}^T$ is defined as

$$\mathcal{N}_{depol.}^T(\rho_i) = p^T \frac{I}{2} + (1 - p^T) U_\theta \rho_i U_\theta^\dagger, \tag{14}$$

where $T$ is the time parameter with $T \in [0, T_{crit.}]$, $U_\theta$ is the polarization rotation due to the relative movement (see Eq. (3)), $p^T$ is the *time-dependent depolarization parameter* of the channel. The time-dependency of the depolarization parameter of the channel is defined as

$$p^T = \begin{cases} p_{depol.}, & \text{if } T \in [0, T_{crit.}] \\ 1, & \text{if } T > T_{crit.} \end{cases} \tag{15}$$

where $p_{dep.}$ is the depolarization parameter of the channel $\mathcal{N}_{depol.}^T$. The depolarization parameter of $\mathcal{N}_{depol.}^T$ takes $p_{dep.}$ for $T \in [0, T_{crit.}]$, and takes it maximum $p_{dep.} = 1$ if $T$ exceeds $T_{crit.}$. For $T > T_{crit.}$ the channel behaves as a completely depolarizing channel

$$\mathcal{N}_{depol.}^T(\rho_i) = \frac{I}{2}. \tag{16}$$

Let assume that Alice uses two orthogonal states $\rho_0 = |0\rangle\langle 0|$ and $\rho_1 = |1\rangle\langle 1|$ for the encoding then the mixed input state of the channel is

$$\rho = \left(\sum_i p_i \rho_i\right) = p_0 \rho_0 + (1 - p_0)\rho_1. \tag{17}$$

After the channel has realized the transformation $\mathcal{N}_{depol.}^T$ on state $\rho$, we will get the following result

$$\begin{aligned}\mathcal{N}_{depol.}^T(\rho) &= \mathcal{N}_{depol.}^T\left[\sum_i p_i \rho_i\right] \\ &= \mathcal{N}_{depol.}^T\left(p_0 U_\theta \rho_0 U_\theta^\dagger + (1-p_0) U_\theta \rho_1 U_\theta^\dagger\right) \\ &= p^T \frac{1}{2} I + (1 - p^T)\left(p_0 U_\theta \rho_0 U_\theta^\dagger + (1-p_0) U_\theta \rho_1 U_\theta^\dagger\right),\end{aligned} \tag{18}$$

where $\theta$ is the polarization rotation that occurs during $T \in [0, T_{crit.}]$.

Geometrically, the map of the time-dependent depolarizing quantum channel shrinks the original Bloch sphere in every direction by $1 - p^T$. First we show the effect of the $\mathcal{N}_{depol.}^T$ on the mixed input pilot state $\rho_{IN}^{Pilot} \neq \frac{1}{2}I$. The $\theta$ rotation of the mixed $\rho_{IN}^{Pilot}$ input pilot state due the relative movement and the map of the time-dependent $\mathcal{N}_{depol.}^T$ depolarizing quantum channel are illustrated in Fig. 3. The shrink of the Bloch vector of state $\rho_{IN}^{Pilot}$ will be relatively small, because the $p^T$ depolarization parameter of channel $\mathcal{N}_{depol.}^T$ depends only on the time parameter $T$.

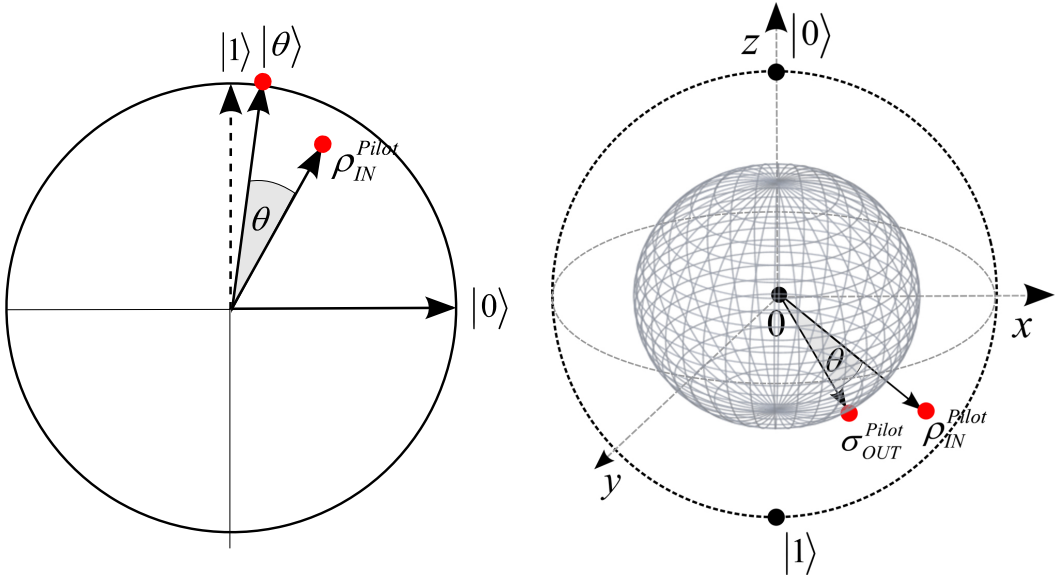

**Figure 3**. (a): The $\rho_{IN}^{Pilot}$ mixed (but non maximally-mixed) input pilot state will be rotated by angle $\theta$ due to the relative movement. The pure state $|\theta\rangle$ is the purification state of the mixed output pilot state $\sigma_{OUT}^{Pilot}$. (b): The channel is a time-dependent depolarizing channel which generates a mixed state $\sigma_{OUT}^{Pilot}$. The channel input pilot state $\rho_{IN}^{Pilot}$ is rotated by angle $\theta$ (grey sphere represents the depolarizing channel map).

The unknown quantum channel $\mathcal{N}$ is described for the $i$-th pure *data* state $|\psi_{A,i}\rangle_{IN}^{Data}$ as follows: it rotates the angle of any input states $|\psi_{A,i}\rangle_{IN}^{Data}$ by angle $\theta$ and shrinks the pure state $|\psi_{A,i}\rangle_{IN}^{Data}$ along to the center of the Bloch sphere. The mixed channel output *data* state $\sigma_{OUT}^{Data}$ will be the following state, according to the map of the time-dependent depolarizing channel:

$$\sigma_{OUT}^{Data} = \mathcal{N}_{depol.}^{T}\left(|\psi_{A,i}\rangle\langle\psi_{A,i}|_{IN}^{Data}\right) = p^T \frac{I}{2} + \left(1-p^T\right) U_\theta |\psi_{A,i}\rangle\langle\psi_{A,i}|_{IN}^{Data} U_\theta^\dagger. \quad (19)$$

In the error correction we will calculate with the purification $|\psi_{A,i}\rangle_{OUT}^{Data}$ of $\sigma_{OUT}^{Data}$, defined as

$$\sigma_{OUT}^{Data} = \mathcal{N}_{depol.}^{T}\left(|\psi_{A,i}\rangle\langle\psi_{A,i}|_{IN}^{Data}\right) = Tr_E |\psi_{A,i}\rangle\langle\psi_{A,i}|_{OUT}^{Data}, \quad (20)$$

where $|\psi_{A,i}\rangle_{OUT}^{Data}$ is the a pure system according to the given $i$-th pure input data state $|\psi_{A,i}\rangle_{IN}^{Data}$. In the decoding process, the purified pilot and data systems $|\theta\rangle$ and $|\psi_{A,i}\rangle_{OUT}^{Data}$ will be used.

## A. Classical Capacity of the Channel

As it was mentioned, the $\mathcal{N}_{depol.}^T$ time-dependent depolarizing channel is a unital channel. Geometrically, this means that the channel maps an identity transformation to an identity transformation, hence $\mathcal{N}_{depol.}^T(I) = I$. We show the channel ellipsoid of the time-dependent depolarizing channel $\mathcal{N}_{depol.}^T$ in Fig. 4.

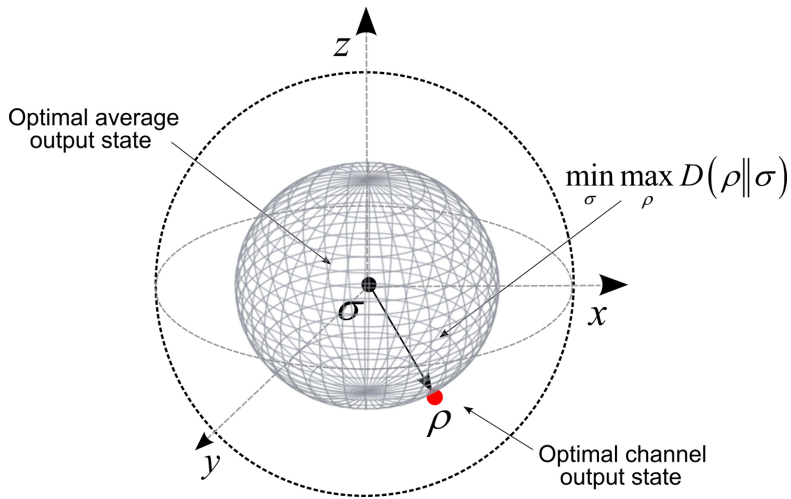

**Figure 4.** The channel ellipsoid of the time-dependent depolarizing quantum channel model. The center of the channel ellipsoid and the smallest quantum informational ball is equal to the center of the of the Bloch sphere.

As depicted in the figure, the HSW capacity [24], [25] of channel $\mathcal{N}_{depol.}^T$ can be expressed by the maximized quantum relative entropy function $\min_{\sigma} \max_{\rho} D(\rho \| \sigma)$ [29], where $D(\cdot \| \cdot)$ is the quantum relative entropy function between quantum states $\rho$ and $\sigma$ is defined as

$$D(\rho \| \sigma) = Tr(\rho \log(\rho)) - Tr(\rho \log(\sigma)) = Tr[\rho(\log(\rho) - \log(\sigma))]. \tag{21}$$

In the definition above, the term $Tr(\rho \log(\sigma))$ is finite only if $\rho \log(\sigma) \geq 0$ for all diagonal matrix elements. If this condition is not satisfied, then $D(\rho \| \sigma)$ could be infinite, since the trace of the second term could go to infinity. The *HSW capacity* of $\mathcal{N}_{depol.}^T$ can be analyzed by the

*optimal average* output state $\sigma$, and the *optimal channel output* state $\rho$ which are generated during $T \in [0, T_{crit.}]$. The average state $\sigma = \sum_k p_k \rho_k$ of the optimal *output* ensembles $\{p_k, \rho_k\}$ is equal to the center of the Bloch sphere. The HSW capacity of $\mathcal{N}_{depol.}^T$ is equal to the quantum relative entropic distance between the optimal output state $\rho$ (which maximizes the channel capacity) and the origin of the Bloch sphere [29], thus for $T \in [0, T_{crit.}]$ we have

$$C(\mathcal{N}_{depol.}^T) = D\left(\rho \middle\| \frac{1}{2} I\right) = 1 - S(\rho), \tag{22}$$

where $S(\cdot \| \cdot)$ is the von Neumann entropy function. The density matrix $\sigma$ must be expressible as a convex combination of $\rho_k$ as $\sigma = \sum_k p_k \rho_k$, which satisfies the min-max criteria of Schumacher and Westmoreland [34], which leads us to HSW capacity of $\mathcal{N}_{depol.}^T$:

$$C(\mathcal{N}_{depol.}^T) = \min_\sigma \max_\rho D(\rho \| \sigma) = \min_\sigma \max_{\rho_k} D\left(\rho_k \middle\| \sigma = \sum_k p_k \rho_k\right), \tag{23}$$

which is equal to

$$C(\mathcal{N}_{depol.}^T) = 1 - \sum_i p_i S\left(\mathcal{N}_{depol.}^T(\rho_i)\right). \tag{24}$$

Furthermore, for $T \in [0, T_{crit.}]$ and $p^T = p_{depol.}$ the following relations hold for the classical capacity of $\mathcal{N}_{depol.}^T$:

$$\begin{aligned}
C(\mathcal{N}_{depol.}^T) &= \min_\sigma \max_\rho D(\rho \| \sigma) = \min_\sigma \max_{\rho_k} D\left(\rho_k \middle\| \sigma = \sum_k p_k \rho_k\right) \\
&= \sum_{\text{all } p_i} p_i D\left(\rho_i \middle\| \frac{1}{2} I\right) = 1 - \sum_i p_i S\left(\mathcal{N}_{depol.}^T(\rho_i)\right) \\
&= \max_{\text{all } p_i, \rho_i} \left[ S\left(\mathcal{N}_{depol.}^T\left(\sum_i p_i \rho_i\right)\right) - \sum_i p_i S\left(\mathcal{N}_{depol.}^T(\rho_i)\right) \right] \\
&= 1 - S\left(p^T \frac{1}{2} I\right) = 1 - H\left(\frac{1}{2} p^T\right) = 1 - H\left(\frac{1}{2} p_{depol.}\right),
\end{aligned} \tag{25}$$

where $H$ is the Shannon entropy, and $p_{depol.}$ is the depolarization parameter of the depolarizing channel.

The classical capacity of the $\mathcal{N}_{depol.}^T$ time-dependent depolarizing channel for $T > T_{crit.}$ is trivially

$$C\left(\mathcal{N}_{depol.}^T\right) = 0. \tag{26}$$

## B. Quantum Capacity of the Channel

The quantum capacity [26-28] of the $\mathcal{N}_{depol.}^T$ time-dependent depolarizing quantum channel for $T \in \left[0, T_{crit.}\right]$ and with depolarization parameter $p_{depol.} \leq 0.25$, (according to the no-cloning bound) can be expressed as

$$Q\left(\mathcal{N}_{depol.}^T\right) \leq 1 - 4p^T = 1 - 4p_{depol.}. \tag{27}$$

Using channel $\mathcal{N}_{depol.}^T$ with $T \in \left[0, T_{crit.}\right]$ it is possible to construct a quantum code with rate at least $R$ that any qubit $|\psi_A\rangle = \alpha|0\rangle + \beta|1\rangle$ in the pilot code can be recovered from the mixed channel output state $\rho$ with fidelity $F = \langle \psi_A | \rho | \psi_A \rangle$. The quantum capacity of channel $\mathcal{N}_{depol.}^T$ for $T > T_{crit.}$ is

$$Q\left(\mathcal{N}_{depol.}^T\right) = 0. \tag{28}$$

The classical and the quantum capacities of the time-dependent $\mathcal{N}_{depol.}^T$ depolarizing channel in function of the time-dependent depolarization parameter $p^T$ are shown in Fig. 5. In $T \in \left[0, T_{crit.}\right]$ the time-dependent depolarizing parameter takes the constant value $p^T = p_{depol.}$. As the $T_{crit.}$ critical time parameter of the channel is exceeded, the depolarizing parameter takes its maximum $p^T = 1$.

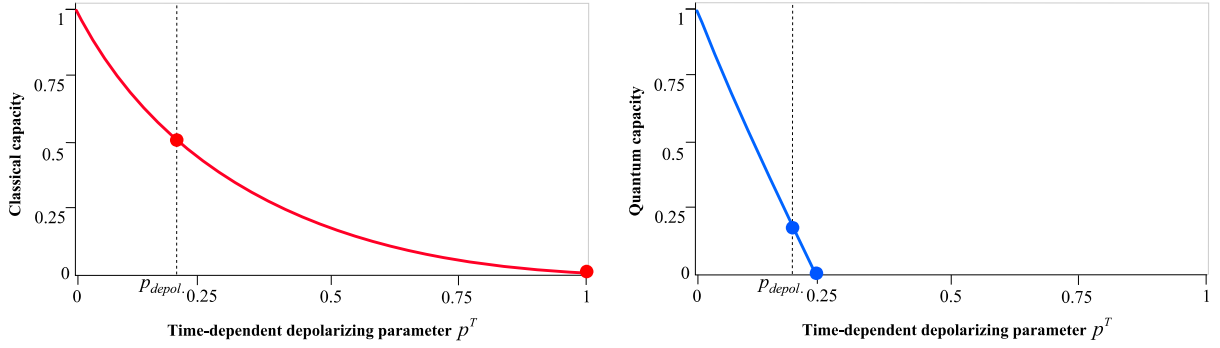

**Figure 5.** The classical and the quantum capacities of the time-dependent quantum channel in function of the time-dependent depolarizing parameter.

The time-dependency of the channel and the static behaviors of the channel capacities as the function of the time parameter $T$ are illustrated in Fig. 6. Positive classical or quantum capacity can be achieved only in time domain $T \in [0, T_{crit.}]$. If $T > T_{crit.}$, the capacities become zero, i.e., $C(\mathcal{N}_{depol.}^T) = Q(\mathcal{N}_{depol.}^T) = 0$. As the $T_{crit.}$ critical time parameter of the channel is exceeded, the channel capacities are cut down to zero.

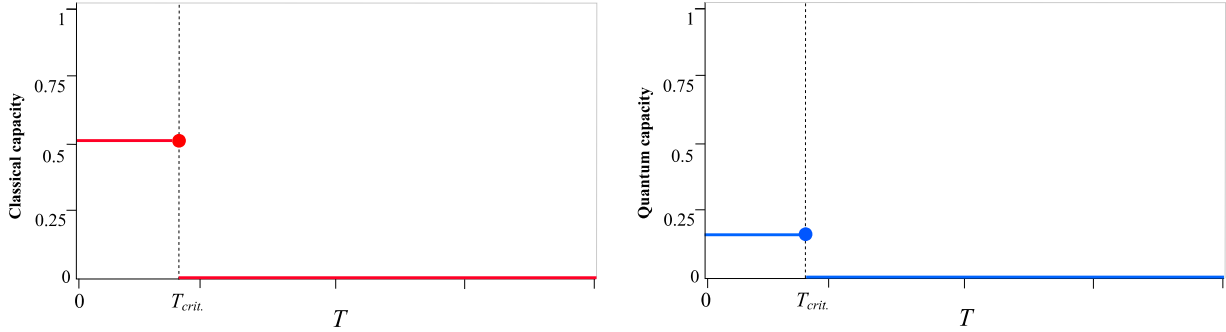

**Figure 6.** The classical (a) and the quantum (b) capacities of the time-dependent depolarizing channel. The capacities of the channel vanish as the time parameter $T$ reaches the critical value of $T_{crit.}$. The channel can transmit classical or quantum information only in the domain $T \in [0, T_{crit.}]$.

Since the capacities of the $\mathcal{N}_{depol.}^T$ time-dependent depolarizing channel decreases as $T \to T_{crit.}$, to achieve high communication rates, low depolarizing parameter $p^T \approx 0$ and sufficiently large $T_{crit.}$ parameters would be desired. As we will show in Section 6, these conditions are almost completely satisfied for a space-earth quantum communication channel, which makes possible to apply the proposed pilot coding scheme with high rates.

## C. Information Transmission over the Time-Dependent Channel

Our error correction method only requires multiple instances of the pilot state, which can be easily achieved by sending $r$ quantum states (For the exact number of $r$ see Section 6. From the $r$ pilot states Bob generates an $l$-length string for the error correction.) with the data qubits over the quantum channel. According to Fig. 1, the quantum channel $\mathcal{N}$ carries out the same transformation (i.e., rotation $\theta$ in the polarization) on all of the pilot and data qubits. The map of the quantum channel is characterized by the self-inverse operator $G$, will be produced on (8), and according to (10)

$$|\theta\rangle = \cos\frac{\theta}{2}|0\rangle + i\sin\frac{\theta}{2}|1\rangle. \tag{29}$$

When receiving, Bob applies a Hadamard transformation on $|\theta\rangle$. Upon reception of the pilot states, Bob can use them to correct the errors of the quantum channel. Using the pilot state $|\theta\rangle$, the polarization rotation can be expressed as Eq. (13). Alice sends her $i$-th quantum state

$$|\psi_{A,i}\rangle = \alpha|0\rangle + \beta|1\rangle, \tag{30}$$

where $\alpha$ and $\beta$ are the probability amplitudes, $\alpha^2 + \beta^2 = 1$, on the quantum channel, which according to (20) is transformed into

$$|d_i\rangle = U_\theta |\psi_{A,i}\rangle. \tag{31}$$

The unitary transformation of the polarization rotation is denoted by $U_\theta$, and defined by Eq. (5). The error correction of data qubits $|d_i\rangle$ will be completed with the help of the pilot qubits $|\theta\rangle$. In the proposed error correction scheme the channel output pilot states are distinguished into two well separable sets. The first set contains the $r$ channel output pilot states, while in the second set we have $l \ll r$ pilot states. In the error correction only the states of the second set are *valuable*, while the elements of the first set are just of required to generate the valuable pilot states, i.e., the elements of the second set. The first set contains the *r*-length string

$$\otimes_{i=1}^{r} |\theta\rangle_i,  \tag{32}$$

while the second *l*-length string contains only the "*2 power*" states

$$\otimes_{i=1}^{l} \left|2^{i-1}\theta\right\rangle.  \tag{33}$$

As we will show in Section 5, only the "2 power" states valuable in the error correcting process. We also will give the exact values of $r$ and $l$. The proposed error-correction method only requires multiple instances of the pilot state (see Fig. 7), which can be easily achieved by sending $r$ quantum states (For the exact number of $r$ see Table 1.) From the $r$ pieces of pilot states Bob generates an *l*-length string for the error-correction, which string contains $l$ valuable pilot quantum states. For the exact connection between the number $r$ of transmitted pilot states and the number $l$ of valuable pilot states see Table 1. Using the pilot state $|\theta\rangle$, the polarization rotation can be expressed as depicted in Eq. (13).

**Input and Output System**

In the communication phase, Alice first feeds to the quantum channel the $r$ pilot (*mixed* but non maximally-mixed) states $\rho_{IN}^{Pilot}$,

$$\rho_{IN}^{Pilot} = \rho_{1,IN}^{Pilot} \otimes \rho_{2,IN}^{Pilot} \otimes \ldots \otimes \rho_{r,IN}^{Pilot},  \tag{34}$$

then the $n$ input data qubits (*pure* states), i.e.,

$$|\psi_A\rangle_{IN}^{Data} = |\psi_{A,1}\rangle \otimes \ldots \otimes |\psi_{A,n}\rangle.  \tag{35}$$

The output system is characterized by the $r$ *purified* channel output pilot states (*pure* states)

$$|\varphi\rangle_{OUT}^{Pilot} = |\theta_1\rangle \otimes |\theta_2\rangle \otimes \ldots \otimes |\theta_r\rangle = \otimes_{i=1}^{r} |\theta_i\rangle  \tag{36}$$

and the $n$ channel *purified* output data states (*pure* states)

$$|\psi_A\rangle_{OUT}^{Data} = |d_1\rangle \otimes |d_2\rangle \otimes \ldots \otimes |d_n\rangle,  \tag{37}$$

where

$$|d_i\rangle = U_\theta |\psi_{A,i}\rangle. \tag{38}$$

From the $r$ channel output (unknown) pilot states, Bob generates an $l$-length pilot string, which contains those valuable pilot states which are required for the error-correction. The elements of this string are the *valuable* "2 power" channel output pilot states (using the pure purification state $|\theta_i\rangle$ of mixed channel output state $\sigma_{OUT,i}^{Pilot} = \mathcal{N}_{depol.}^T\left(\rho_{i,IN}^{Pilot}\right) = Tr_E |\theta_i\rangle\langle\theta_i|$ )

$$\begin{aligned}|\theta\rangle^{Pilot} &= \otimes_{i=1}^{l}\left|2^{i-1}\theta\right\rangle = |\theta_1\rangle \otimes |\theta_2\rangle \otimes \ldots \otimes |\theta_l\rangle \\ &= \left\{|\theta\rangle, |2\theta\rangle, |4\theta\rangle, \ldots, \left|2^{l-1}\theta\right\rangle\right\}.\end{aligned} \tag{39}$$

The quantum channel is represented by $U_\theta$, and defined by Eq. (5). The error-correction of output data qubits $|d_i\rangle$ will be carried out by means of the received purified pilot qubits $|\theta\rangle$. The determination of the ratio between the data qubit states and pilot states depends on the physical properties (time parameter $T$) of the quantum channel [16]. The number $r$ of pilot qubits can be chosen to be several orders of magnitude lower than the number $n$ of data qubits (see the theorems and proofs of Section 5).

The *mixed* pilot qubits $\rho_{IN}^{Pilot}$ along with the *pure* data qubits $|\psi_A\rangle_{IN}^{Data}$ are transmitted from Alice to Bob, as shown in Fig. 7. The purified output systems are denoted by $|\varphi\rangle_{OUT}^{Pilot}$ and $|\psi_A\rangle_{OUT}^{Data}$. In the transmission phase, Alice sends the mixed pilot qubits and the data qubits. This *time-dependency* in the encoding and decoding process makes it possible to Bob to distinguish between the two sets of channel outputs $\sigma_{OUT}^{Pilot}$ and $\sigma_{OUT}^{Data}$, i.e., the purified states $|\theta\rangle$ and $|\psi_A\rangle_{OUT}^{Data}$ can be identified unambiguously. Alice's initial state is $|\psi_{A,i}\rangle$, the error correction transformation is the inverse of the channel's transformation on $|d_i\rangle$, i.e., considering a unitary transformation, it is the adjugate operator $U_\theta^\dagger$.

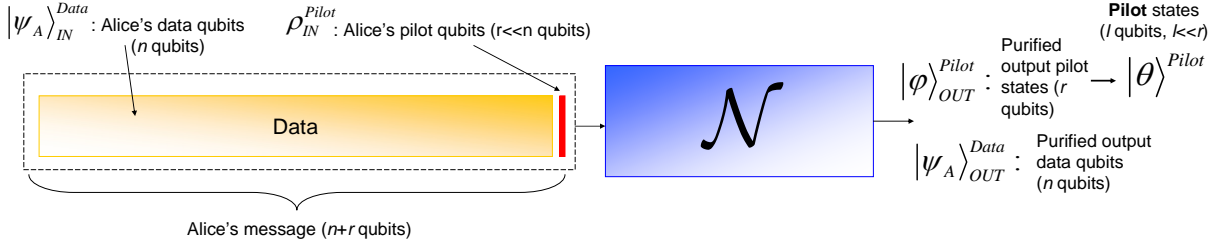

**Figure 7.** Alice sends the data qubits and a very small number of pilot states. The pilot qubits will store the unknown transformation of the quantum channel. The data qubits will be corrected with the help of pilot qubits.

**Remark 1**. The data qubits can carry both *classical* and *quantum* information. In case of classical information the data qubits are referred as *known* data states, for quantum information as *unknown* states. For *classical* information, the data block can be constructed by a simple classical *repetition code*. In case of the transmission of *quantum* information, the data block is filled with *random* qubits.

## V. Theorems and Proofs

In this section we present the theorems and the proofs and prove the correctness of the proposed pilot quantum encoding scheme (*Note*: The results will be demonstrated for qubit inputs ($d=2$ dimensional systems) and qubit channels.)

### A. Determination of the Presence of Noise

**Theorem 1**. *The presence of noise can be determined from the state of the purified pilot state $|\theta\rangle$.*

*Proof.* Bob constructs a simple quantum circuit, which can solve this problem, but only in a probabilistic way. In Fig. 8, Bob's quantum circuit is shown, constructed to distinguish between the purified channel output pilot state $|\theta\rangle$ and the pure reference pilot state $|\varphi\rangle_{\text{REF.}}^{Pilot}$ (known a-priori by Bob). The verifier circuit contains only Hadamard-gates and a *controlled-SWAP*-gate. (The SWAP gate can change the two inputs on its output.).

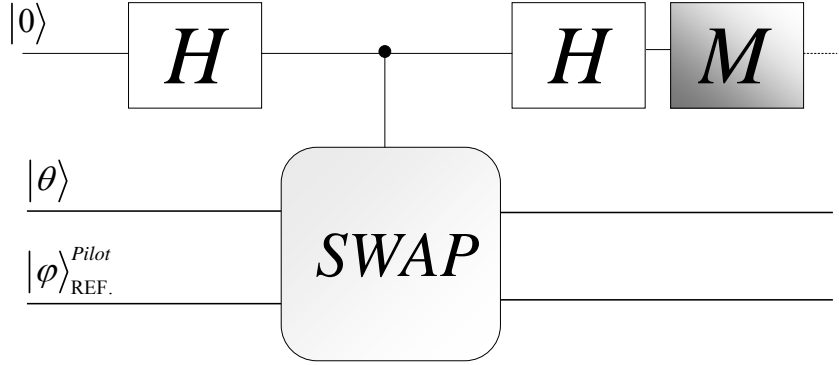

**Figure 8.** Bob's verifier circuit to determine the presence of noise in the quantum channel. The purified pilot state $|\varphi\rangle$ will be compared with the reference pilot state $|\varphi\rangle_{\text{REF.}}^{Pilot}$, which state is a-priori known by Bob.

On the output of the circuit, Bob will measure a 0 or 1. From this measurement result he will be able to determine whether the channel is ideal or noisy. Bob's zero output can be generated by the following steps of the quantum circuit:

$$\begin{aligned}
&|0\rangle \otimes |\theta\rangle \otimes |\varphi\rangle_{\text{REF.}}^{Pilot} \\
&\overset{H}{\mapsto} \frac{1}{\sqrt{2}} \Big( |0\rangle \otimes |\theta\rangle \otimes |\varphi\rangle_{\text{REF.}}^{Pilot} + |1\rangle \otimes |\theta\rangle \otimes |\varphi\rangle_{\text{REF.}}^{Pilot} \Big) \\
&= \frac{1}{\sqrt{2}} \Big( |0\rangle \otimes |\theta\rangle \otimes |\varphi\rangle_{\text{REF.}}^{Pilot} + |1\rangle \otimes |\theta\rangle \otimes |\varphi\rangle_{\text{REF.}}^{Pilot} + |0\rangle \otimes |\varphi\rangle_{\text{REF.}}^{Pilot} \otimes |\theta\rangle - |1\rangle \otimes |\varphi\rangle_{\text{REF.}}^{Pilot} \otimes |\theta\rangle \Big) \\
&\overset{SWAP,\ H}{\mapsto} |0\rangle \otimes \frac{1}{2} \Big( |\theta\rangle \otimes |\varphi\rangle_{\text{REF.}}^{Pilot} + |\varphi\rangle_{\text{REF.}}^{Pilot} \otimes |\theta\rangle \Big) + |1\rangle \otimes \frac{1}{2} \Big( |\theta\rangle \otimes |\varphi\rangle_{\text{REF.}}^{Pilot} - |\varphi\rangle_{\text{REF.}}^{Pilot} \otimes |\theta\rangle \Big).
\end{aligned}$$
(40)

From this result, the output probability of the 0 outcome can be expressed as follows:

$$\begin{aligned}
\Pr[0] &= \frac{1}{4} \Big( \langle \theta, \varphi_{\text{REF.}}^{Pilot} | + \langle \varphi_{\text{REF.}}^{Pilot}, \theta | \Big) \Big( |\theta, \varphi_{\text{REF.}}^{Pilot}\rangle + |\varphi_{\text{REF.}}^{Pilot}, \theta\rangle \Big) \\
&= \frac{1}{4} \Big( 2 + \langle \varphi_{\text{REF.}}^{Pilot}, \theta | \theta, \varphi_{\text{REF.}}^{Pilot} \rangle + \langle \theta, \varphi_{\text{REF.}}^{Pilot} | \varphi_{\text{REF.}}^{Pilot}, \theta \rangle \Big) \\
&= \frac{1}{2} + \frac{1}{2} \big| \langle \varphi_{\text{REF.}}^{Pilot} | \theta \rangle \big|^2.
\end{aligned}$$
(41)

Bob would like to measure 0, if the two states are equal, i.e., their inner product is $\langle \theta | \varphi_{\text{REF.}}^{Pilot} \rangle \approx 1$, and he hopes measuring 1, if the two states are different (nearly orthogonal to each other), i.e., $\langle \theta | \varphi_{\text{REF.}}^{Pilot} \rangle \approx 0$. The probability of 0 outcome is

$$\Pr\left[0 \middle| \langle \theta | \varphi_{\text{REF.}}^{Pilot} \rangle \right] = \frac{1}{2} + \frac{\langle \theta | \varphi_{\text{REF.}}^{Pilot} \rangle}{2} \Rightarrow \Pr\left[0 \middle| \langle \theta | \varphi_{\text{REF.}}^{Pilot} \rangle \approx 1\right] = \frac{1}{2} + \frac{\overbrace{\langle \theta | \varphi_{\text{REF.}}^{Pilot} \rangle}^{\approx 1}}{2} \approx 1, \quad (42)$$

while measuring 1 has probability

$$\Pr\left[1 \middle| \langle \theta | \varphi_{\text{REF.}}^{Pilot} \rangle \approx 0\right] = \frac{1}{2} - \frac{\overbrace{\langle \theta | \varphi_{\text{REF.}}^{Pilot} \rangle}^{\approx 0}}{2} \approx \frac{1}{2}. \quad (43)$$

As can be seen, when $\langle \theta | \varphi_{\text{REF.}}^{Pilot} \rangle \approx 1$, Bob can almost always generate the correct answer, but in case of $\langle \theta | \varphi_{\text{REF.}}^{Pilot} \rangle \approx 0$, he may incorrectly conclude that the two systems equal when this is not, in fact, the case.

**Proposition 1**. *Bob can determine the presence of the noise from states $|\theta\rangle$ and $|\varphi\rangle_{\text{REF.}}^{Pilot}$ with arbitrary low error.*

If Alice sends to Bob some copies from $|\theta\rangle$ and the reference pilot state $|\varphi\rangle_{\text{REF.}}^{Pilot}$, and Bob can repeat the test multiple times, then Bob can increase the probability of success. However, this step decreases the efficiency of the quantum circuit, but the error probability of the network also can be decreased.

The outcome probabilities can be used to derive the error probabilities for the case of $|\theta\rangle = |\varphi_{\text{REF.}}^{Pilot}\rangle$ and $|\theta\rangle \neq |\varphi_{\text{REF.}}^{Pilot}\rangle$. In the first case, the quantum circuit of Bob gives a correct output with error probability

$$p_{error}^{|\theta\rangle=|\varphi_{\text{REF.}}^{Pilot}\rangle} = 1 - \left[\frac{1}{2} + \frac{(1-\delta)}{2}\right] = \frac{1}{2} - \frac{(1-\delta)}{2} = 0, \tag{44}$$

where $\delta = 1 - \langle\theta|\varphi_{\text{REF.}}^{Pilot}\rangle$, which is trivially zero if $|\theta\rangle$ and $|\varphi\rangle_{\text{REF.}}^{Pilot}$ are equal. On the other hand, in the second case, the error probability will be higher, thus the worst-case error probability for the $|\theta\rangle \neq |\varphi_{\text{REF.}}^{Pilot}\rangle$ input is

$$p_{error}^{|\theta\rangle\neq|\varphi_{\text{REF.}}^{Pilot}\rangle} = 1 - \left[\frac{1}{2} - \frac{(1-\delta)}{2}\right] = \frac{1}{2} + \frac{(1-\delta)}{2}. \tag{45}$$

But, Bob can decrease this error probability to an arbitrary $\varepsilon > 0$ error. To reach this arbitrary $\varepsilon > 0$ error in the case of $|\theta\rangle \neq |\varphi_{\text{REF.}}^{Pilot}\rangle$, Bob has to repeat $k$-times the test with the constructed quantum circuit, where

$$k \in \mathcal{O}\left(\log_2(1/\varepsilon)\right). \tag{46}$$

After $k$-iterations, the error probability reduces to

$$p_{error}^{|\theta\rangle\neq|\varphi_{\text{REF.}}^{Pilot}\rangle} = \left(1 - \left[\frac{1}{2} - \frac{(1-\delta)}{2}\right]\right)^k = \left(\frac{1}{2} + \frac{(1-\delta)}{2}\right)^k, \tag{47}$$

which for $(1-\delta) \to 0$, hence $|\langle\theta|\varphi_{\text{REF.}}^{Pilot}\rangle| \leq 1 - \delta$, which results in a probability of making an error in the case of logical 0 of

$$p_{error} = \frac{1}{2^k}. \tag{48}$$

It proves that Bob can determine the question whether $|\theta\rangle$ and $|\varphi\rangle_{\text{REF.}}^{Pilot}$ are equal or not, with vanishing error probability.

These results conclude the proof of Theorem 1.

∎

## B. The Error Capturing Process

The main result on the fundament of the proposed error correction scheme is summarized in Theorem 2.

**Theorem 2.** *The $U_\theta$ operator of the unknown polarization rotation of the quantum channel can be stored in the pilot quantum state $|\theta\rangle$, where $|\theta\rangle$ is the purification state of the mixed channel output pilot state $\sigma_{OUT}^{Pilot} = \mathcal{N}_{depol.}^{T}\left(\rho_{IN}^{Pilot}\right) = Tr_E |\theta\rangle\langle\theta|$, with the input condition $\rho_{IN}^{Pilot} \neq \frac{1}{2}I$.*

*Proof.* The transformation $U_\theta$ of the channel will be stored in the *pilot quantum* state

$$|\theta\rangle = \cos\frac{\theta}{2}|0\rangle + i\sin\frac{\theta}{2}|1\rangle, \qquad (49)$$

thus $U_\theta$, the polarization rotation transformation of the given angle $\theta$ can be rewritten as

$$U_\theta = e^{i\frac{\theta}{2}G} = e^{i\frac{\theta}{2}\begin{bmatrix}1 & 0\\ 0 & -1\end{bmatrix}}, \qquad (50)$$

where $G$ is the generator matrix, and (50) which can be further evaluated as

$$\begin{aligned}
U_\theta &= \cos\frac{\theta}{2}I + i\sin\frac{\theta}{2}\begin{bmatrix}1 & 0\\ 0 & -1\end{bmatrix}\\
&= \cos\frac{\theta}{2}I + i\sin\frac{\theta}{2}Z\\
&= \cos\frac{\theta}{2}|0\rangle\langle 0| + \cos\frac{\theta}{2}|1\rangle\langle 1| + i\sin\frac{\theta}{2}|0\rangle\langle 0| - i\sin\frac{\theta}{2}|1\rangle\langle 1|\\
&= \left(\cos\frac{\theta}{2} + i\sin\frac{\theta}{2}\right)|0\rangle\langle 0| + \left(\cos\frac{\theta}{2} - i\sin\frac{\theta}{2}\right)|1\rangle\langle 1|\\
&= \begin{pmatrix}\left(\cos\frac{\theta}{2} + i\sin\frac{\theta}{2}\right) & 0\\ 0 & \left(\cos\frac{\theta}{2} - i\sin\frac{\theta}{2}\right)\end{pmatrix}\\
&= e^{i\frac{\theta}{2}\begin{bmatrix}1 & 0\\ 0 & -1\end{bmatrix}},
\end{aligned} \qquad (51)$$

i.e., the generator matrix of $U_\theta$ is

$$G = \begin{bmatrix} 1 & 0 \\ 0 & -1 \end{bmatrix}. \tag{52}$$

These results mean that the transformation $G=Z$ is mapped onto the state $|1\rangle$, while - provided the channel has no error and hence realizes an identity transformation $I$ - is mapped onto $|0\rangle$. From these result follows the polarization rotation angle $\theta$ can be stored in the pilot state $|\theta\rangle$, and the transformation $U_\theta$ can be restored from the pilot state $|\theta\rangle$, since the unknown transformation is mapped onto the state $|1\rangle$. *Note*: The generator matrix $G$ is known by both Alice and Bob.

These results conclude the proof of Theorem 2.

∎

The main result on the single data qubit error-correction is summarized in Theorem 3.

## C. The Error Correction Process

The correction of an arbitrary *data* state $|d_i\rangle$ can be achieved by the $|\theta\rangle$ pilot state and elementary quantum gates. The result for the single qubit case is stated as follows.

**Theorem 3.** *The unknown pilot state $|\theta\rangle$ can be used to correct the data state $|d_i\rangle$, of $|\psi_A\rangle_{OUT}^{Data} = |d_0\rangle \otimes |d_1\rangle \otimes ... \otimes |d_{n-1}\rangle$ where $|\psi_A\rangle_{OUT}^{Data}$ is the pure output data system, given by $\sigma_{OUT}^{Data} = \mathcal{N}_{depol.}^T \left( |\psi_A\rangle\langle\psi_A|_{IN}^{Data} \right) = Tr_E |\psi_A\rangle\langle\psi_A|_{OUT}^{Data}$. The error-correction requires only a Hadamard gate and a CNOT gate with $|0\rangle$ control at the receiver.*

*Proof.* Bob does not need to know anything about the channel's unitary transformation $U_\theta$, nor about the data qubit $|d_i\rangle$. Bob cannot exactly determine the received state since he does not know the angle of the error $\theta_i$. In this phase, Bob cannot be sure whether or not the *i*-th quantum state $|d_i\rangle$ is identical to the originally sent state, $|\psi_i\rangle$. We construct a quantum circuit, which uses the *pilot* state and the data states to complete the error correction. As will be explained, we need the inverse transformation stored by the pilot state (and hence $U_\theta^\dagger$, the adjugate of the channel transformation $U_\theta$).

The error-correction method consists of a *control qubit* $|0\rangle$, which corresponds to the modified qubit $|d\rangle$, and a *target qubit*, which is equal to the *error-correction* pilot state $|\theta\rangle$. To correct state $|d\rangle$ to $\psi_A$, Bob uses a simple Hadamard and CNOT transformation with $|0\rangle$ *control*, thus the state $|d\rangle$ is transformed into

$$
\begin{aligned}
|d\rangle \otimes H|\theta\rangle &\to \\
&\frac{1}{\sqrt{2}}\left(U_\theta^\dagger|d\rangle \otimes |0\rangle + U_\theta|d\rangle \otimes |1\rangle\right) \\
&= \frac{1}{\sqrt{2}}\left(U_\theta^\dagger U_\theta\left(|\psi_A\rangle\right) \otimes |0\rangle + U_\theta U_\theta\left(|\psi_A\rangle\right) \otimes |1\rangle\right) = \\
&= \frac{1}{\sqrt{2}}\left(|\psi_A\rangle \otimes |0\rangle + U_\theta U_\theta\left(|\psi_A\rangle\right) \otimes |1\rangle\right),
\end{aligned}
\tag{53}
$$

and therefore a projective measurement in the $\{|0\rangle,|1\rangle\}$ basis of the correction-state $|\theta\rangle$ will make the modified qubit $|d\rangle$ either collapse into the desired, error-corrected state $U_\theta^\dagger|d\rangle$ or into the incorrect state $U_\theta|d\rangle$ [12,13]. That projective measurement will make the damaged state either collapse into the *wrong* state (in the case of measurement outcome $|1\rangle$)

$$U_\theta|d\rangle = U_\theta U_\theta|\psi_A\rangle \tag{54}$$

or into the *right* state (in the case of measurement outcome $|0\rangle$)

$$U_\theta^\dagger|d\rangle = U_\theta^\dagger U_\theta|\psi_A\rangle = |\psi_A\rangle, \tag{55}$$

with each outcome having a probability of 1/2. Therefore, Bob applies the gate of Fig. 9 to prepare the *right state* $U_\theta^\dagger|d\rangle$ or the *wrong state* $U_\theta|d\rangle$ with an equal probability of 1/2.

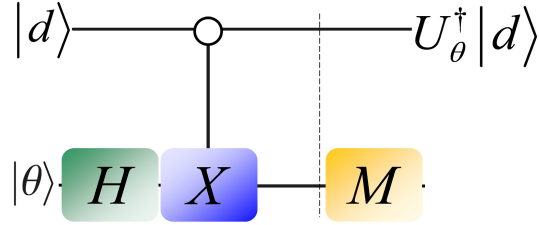

**Figure 9.** The error correction of a damaged qubit with a single-qubit pilot state. The pilot state is one qubit. The gate is controlled by zero.

The CNOT transformation is controlled by the $|0\rangle$, instead of the $|1\rangle$ control qubit. The measurement will be made on the pilot state, which transforms the data qubit into the desired state with a given probability. The error-correction quantum circuit is probabilistic, however with the help of pilot states the success probability can be increased to be arbitrarily high [12,13].

**Proposition 2**. *The pilot error-correction defines a probabilistic process but the success probability can be increased to arbitrarily high.*

The working mechanism of Bob's error correction gate uses the unknown pilot state $|\theta\rangle = \cos\frac{\theta}{2}|0\rangle + i\sin\frac{\theta}{2}|1\rangle$, which stores the transformation of the channel - and the unknown data state

$$|d\rangle = a|0\rangle + b|1\rangle \tag{56}$$

which is a damaged state (i.e., its polarization is rotated). The error correction only requires a CNOT gate (and a Hadamard gate) with $|0\rangle$ control on Bob's side. On the other hand, it is currently a probabilistic mechanism, although as we will show the success probability can be increased arbitrarily high. If Bob has an *l*-length qubit string

$$|\theta\rangle^{Pilot} = \otimes_{i=1}^{l} |2^{i-1}\theta\rangle \tag{57}$$

to decode the damaged state $|d\rangle$, Bob's failure probability will be only

$$\varepsilon = (1/2)^l. \tag{58}$$

The probability of wrong decoding decreases exponentially with the size of $|\theta\rangle$, the length of the error-correction string is denoted by $l$. Bob takes the damaged qubit $|d\rangle$ as the control-bit, and takes error correction qubit states $\otimes_{i=1}^{l} |2^{i-1}\theta\rangle$ as the target, therefore Bob evolves the transformation of

$$|d\rangle \otimes |\theta\rangle^{Pilot} = |d\rangle \otimes_{i=1}^{l} |2^{i-1}\theta\rangle \tag{59}$$

into

$$\frac{1}{\sqrt{2^l}}\left(\sqrt{2^l - 1} U_\theta^{(2^l-1)\dagger} |d\rangle \otimes |right\rangle + U_\theta |d\rangle \otimes |wrong\rangle\right), \tag{60}$$

where $\langle right|wrong\rangle = 0$. The gate for improved decoding is shown in Fig. 10.

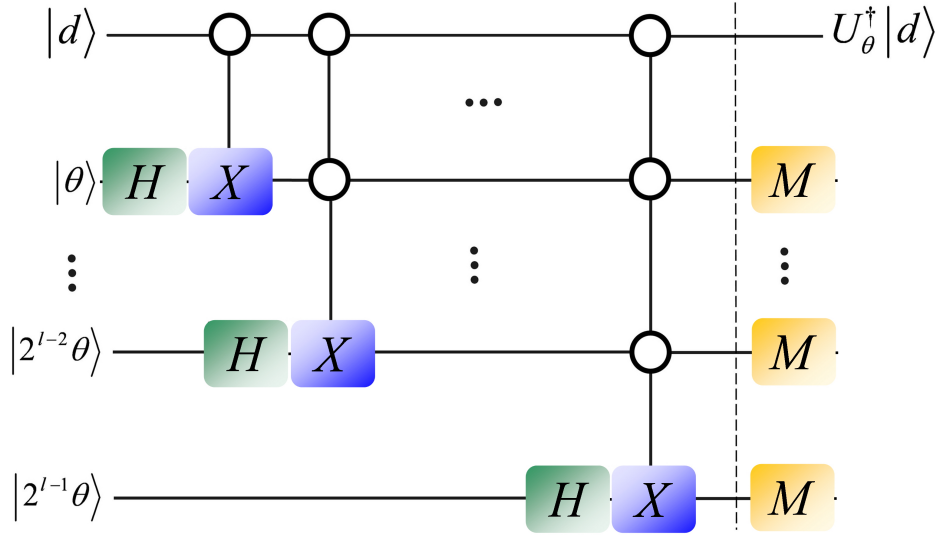

**Figure 10.** The correction of a damaged data state with a multi-qubit pilot state. The error correction angle is stored in an *l*-length quantum string.

As follows, if Bob knows the pilot states $|\theta\rangle^{Pilot} = \otimes_{i=1}^{l}|2^{i-1}\theta\rangle = \{|\theta\rangle,|2\theta\rangle,|4\theta\rangle,...,|2^{l-1}\theta\rangle\}$, then he can correct the damaged qubit with a success probability of

$$p = 1 - (1/2)^l. \tag{61}$$

In this case the right transformation $U_\theta^\dagger$ succeeds with exponentially increasing probability $p = 1 - (1/2)^l$ as the number $l$ of the pilot states increases linearly, and exhibits an exponentially decreasing error probability $\varepsilon = (1/2)^l$ [12,13,16]. Since Bob has a one-qubit length state for correcting the damaged state, Bob fails to perform

$$U_\theta^\dagger \left( U_\theta |\psi_A\rangle \right) = |\psi_A\rangle \tag{62}$$

with a probability $p_1 = 1/2$. The single data qubit error-correction can be achieved as follows:

$$\begin{aligned}
&CNOT\left(|d\rangle \otimes H|\theta\rangle\right) \\
&= (a|0\rangle + b|1\rangle) H\left(\cos\frac{\theta}{2}|0\rangle + i\sin\frac{\theta}{2}|1\rangle\right) \\
&= (a|0\rangle + b|1\rangle)\left(\frac{1}{\sqrt{2}}\left[\left(\cos\frac{\theta}{2} + i\sin\frac{\theta}{2}\right)|0\rangle + \left(\cos\frac{\theta}{2} - i\sin\frac{\theta}{2}\right)|1\rangle\right]\right) \\
&= \frac{1}{\sqrt{2}}\begin{bmatrix} \left[a\left(\cos\frac{\theta}{2} - i\sin\frac{\theta}{2}\right)|0\rangle + b\left(\cos\frac{\theta}{2} + i\sin\frac{\theta}{2}\right)|1\rangle\right]|0\rangle \\ +\left[a\left(\cos\frac{\theta}{2} + i\sin\frac{\theta}{2}\right)|0\rangle + b\left(\cos\frac{\theta}{2} - i\sin\frac{\theta}{2}\right)|1\rangle\right]|1\rangle \end{bmatrix} \\
&= \frac{1}{\sqrt{2}}\begin{bmatrix} \left[\left(\cos\frac{\theta}{2} - i\sin\frac{\theta}{2}\right)|0\rangle\langle 0| + \left(\cos\frac{\theta}{2} + i\sin\frac{\theta}{2}\right)|1\rangle\langle 1|\right]|d\rangle|0\rangle \\ +\left[\left(\cos\frac{\theta}{2} + i\sin\frac{\theta}{2}\right)|0\rangle\langle 0| + \left(\cos\frac{\theta}{2} - i\sin\frac{\theta}{2}\right)|1\rangle\langle 1|\right]|d\rangle|1\rangle \end{bmatrix} \\
&= \frac{1}{\sqrt{2}}\begin{pmatrix} \begin{bmatrix} \cos\frac{\theta}{2} - i\sin\frac{\theta}{2} & 0 \\ 0 & \cos\frac{\theta}{2} + i\sin\frac{\theta}{2} \end{bmatrix}|d\rangle|0\rangle \\ + \begin{bmatrix} \cos\frac{\theta}{2} + i\sin\frac{\theta}{2} & 0 \\ 0 & \cos\frac{\theta}{2} - i\sin\frac{\theta}{2} \end{bmatrix}|d\rangle|1\rangle \end{pmatrix} \\
&= \frac{1}{\sqrt{2}}\left(U_\theta^\dagger|d\rangle|0\rangle + [U_\theta]|d\rangle|1\rangle\right),
\end{aligned} \tag{63}$$

hence Bob can correct any error of the unknown quantum state without the explicit knowledge of the quantum channel. On the other hand, it is currently a probabilistic mechanism, although as we will show the success probability can be increased arbitrarily high as will be provided in Theorem 5. The error correction can be extended to an arbitrarily high number of input quantum data states, instead of just a single qubit as it is summarized in Theorem 6.

These results conclude the proof of Theorem 3.

∎

The result on the generation process of new pilot states for the polarization compensation is summarized in Theorem 4. For the description of the multi-qubit error correction process, see Theorem 6.

## D. The Pilot String Generation Process

The most critical part of the proposed pilot quantum error correction protocol is the pilot string generation. Here we prove that the pilot string generation makes it possible to achieve the error correction with arbitrary high probability; however the channel output states are unknown states. As Theorem 4 shows the unknown *valuable* pilot states can be generated from the unknown channel output pilot states, using elementary quantum gates.

**Theorem 4.** *The l-length string $|\theta\rangle^{Pilot} = \otimes_{i=1}^{l} |2^{i-1}\theta\rangle$ of the unknown valuable pilot states $\{|\theta\rangle, |2\theta\rangle, |4\theta\rangle, \ldots, |2^{l-1}\theta\rangle\}$ can be generated using the r-length string $|\varphi\rangle_{OUT}^{Pilot} = \otimes_{i=1}^{r} |\theta\rangle_i$ of the unknown channel output pilot states. Only the $|\theta\rangle^{Pilot} = \otimes_{i=1}^{l} |2^{i-1}\theta\rangle$ valuable "2 power" pilot states can be used in the error correcting.*

*Proof.* In the proof of Theorem 3 we have demonstrated that Bob can correct any arbitrary length set of data quantum states without knowledge of the transformation of the quantum channel. The generation of these states is performed by the same quantum circuit as used for error-correction, however in this case Bob will use the CNOT with control $|1\rangle$ and will use the pilot states as both the control and the target, as depicted in Fig. 11.

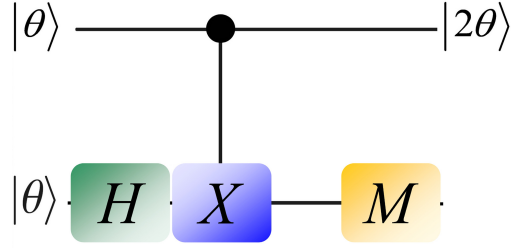

**Figure 11.** The generation of pilot states for high success probability error correction.

The outcome of the circuit shown in Fig. 11 is

$$
\begin{aligned}
&CNOT\left(|\theta\rangle \otimes H|\theta\rangle\right) \\
&= \left(\cos\frac{\theta}{2}|0\rangle + i\sin\frac{\theta}{2}|1\rangle\right) H \left(\cos\frac{\theta}{2}|0\rangle + i\sin\frac{\theta}{2}|1\rangle\right) \\
&= \left(\cos\frac{\theta}{2}|0\rangle + i\sin\frac{\theta}{2}|1\rangle\right) \otimes \frac{1}{\sqrt{2}} \left[\begin{array}{l}\left(\cos\frac{\theta}{2} + i\sin\frac{\theta}{2}\right)|0\rangle \\ +\left(\cos\frac{\theta}{2} - i\sin\frac{\theta}{2}\right)|1\rangle\end{array}\right] \\
&= \frac{1}{\sqrt{2}} \left(\begin{array}{l}\cos\frac{\theta}{2}\left(\cos\frac{\theta}{2} + i\sin\frac{\theta}{2}\right)|00\rangle + \cos\frac{\theta}{2}\left(\cos\frac{\theta}{2} - i\sin\frac{\theta}{2}\right)|01\rangle \\ +i\sin\frac{\theta}{2}\left(\cos\frac{\theta}{2} + i\sin\frac{\theta}{2}\right)|10\rangle + i\sin\frac{\theta}{2}\left(\cos\frac{\theta}{2} - i\sin\frac{\theta}{2}\right)|11\rangle\end{array}\right) \\
&= \frac{1}{\sqrt{2}} \left(\begin{array}{l}\cos\frac{\theta}{2}\left(\cos\frac{\theta}{2} + i\sin\frac{\theta}{2}\right)|00\rangle + \cos\frac{\theta}{2}\left(\cos\frac{\theta}{2} - i\sin\frac{\theta}{2}\right)|01\rangle \\ +i\sin\frac{\theta}{2}\left(\cos\frac{\theta}{2} + i\sin\frac{\theta}{2}\right)|11\rangle + i\sin\frac{\theta}{2}\left(\cos\frac{\theta}{2} - i\sin\frac{\theta}{2}\right)|10\rangle\end{array}\right) \\
&= \frac{1}{\sqrt{2}} \left(\begin{array}{l}\left[\cos\frac{\theta}{2}\left(\cos\frac{\theta}{2} + i\sin\frac{\theta}{2}\right)|0\rangle + i\sin\frac{\theta}{2}\left(\cos\frac{\theta}{2} - i\sin\frac{\theta}{2}\right)|1\rangle\right]|0\rangle \\ +\left[\cos\frac{\theta}{2}\left(\cos\frac{\theta}{2} - i\sin\frac{\theta}{2}\right)|0\rangle + i\sin\frac{\theta}{2}\left(\cos\frac{\theta}{2} - i\sin\frac{\theta}{2}\right)|1\rangle\right]|1\rangle\end{array}\right)
\end{aligned}
$$

$$
\begin{aligned}
&= \frac{1}{\sqrt{2}} \left( \begin{bmatrix} \left[\left(\cos\frac{\theta}{2} + i\sin\frac{\theta}{2}\right)|0\rangle\langle 0| + \left(\cos\frac{\theta}{2} - i\sin\frac{\theta}{2}\right)|1\rangle\langle 1|\right]|\theta\rangle|0\rangle \\ + \left[\left(\cos\frac{\theta}{2} - i\sin\frac{\theta}{2}\right)|0\rangle\langle 0| + \left(\cos\frac{\theta}{2} + i\sin\frac{\theta}{2}\right)|1\rangle\langle 1|\right]|\theta\rangle|1\rangle \end{bmatrix} \right) \\
&= \frac{1}{\sqrt{2}} \left( \begin{bmatrix} \cos\frac{\theta}{2} + i\sin\frac{\theta}{2} & 0 \\ 0 & \cos\frac{\theta}{2} - i\sin\frac{\theta}{2} \end{bmatrix} |\theta\rangle|0\rangle \\ + \begin{bmatrix} \cos\frac{\theta}{2} - i\sin\frac{\theta}{2} & 0 \\ 0 & \cos\frac{\theta}{2} + i\sin\frac{\theta}{2} \end{bmatrix} |\theta\rangle|1\rangle \right) \\
&= \frac{1}{\sqrt{2}} \left( |2\theta\rangle|0\rangle + \left[U_\theta^\dagger\right]|\theta\rangle|1\rangle \right) \\
&= \frac{1}{\sqrt{2}} \left( \left[U_\theta\right]|\theta\rangle|0\rangle + \left[U_\theta^\dagger\right]|\theta\rangle|1\rangle \right) \\
&= \frac{1}{\sqrt{2}} \left( |2\theta\rangle|0\rangle + \left[U_\theta^\dagger\right]|\theta\rangle|1\rangle \right),
\end{aligned}
$$

hence, Bob obtains the next required pilot state $|2\theta\rangle$. This process can be continued, and the new pilot states can be used as control qubits to increase the success probability of the next rotation. The new pilot states can be used as control qubits to increase the success probability of the next rotation.

In Fig. 12 we illustrate the generation process of the 3-length pilot qubit string $|\theta\rangle^{Pilot} = \{|\theta\rangle, |2\theta\rangle, |4\theta\rangle\}$. In the generation process all states and the whole set $|\varphi\rangle_{OUT}^{Pilot} = \otimes_{i=1}^{r} |\theta\rangle_i$ from which the state will be generated are completely unknown.

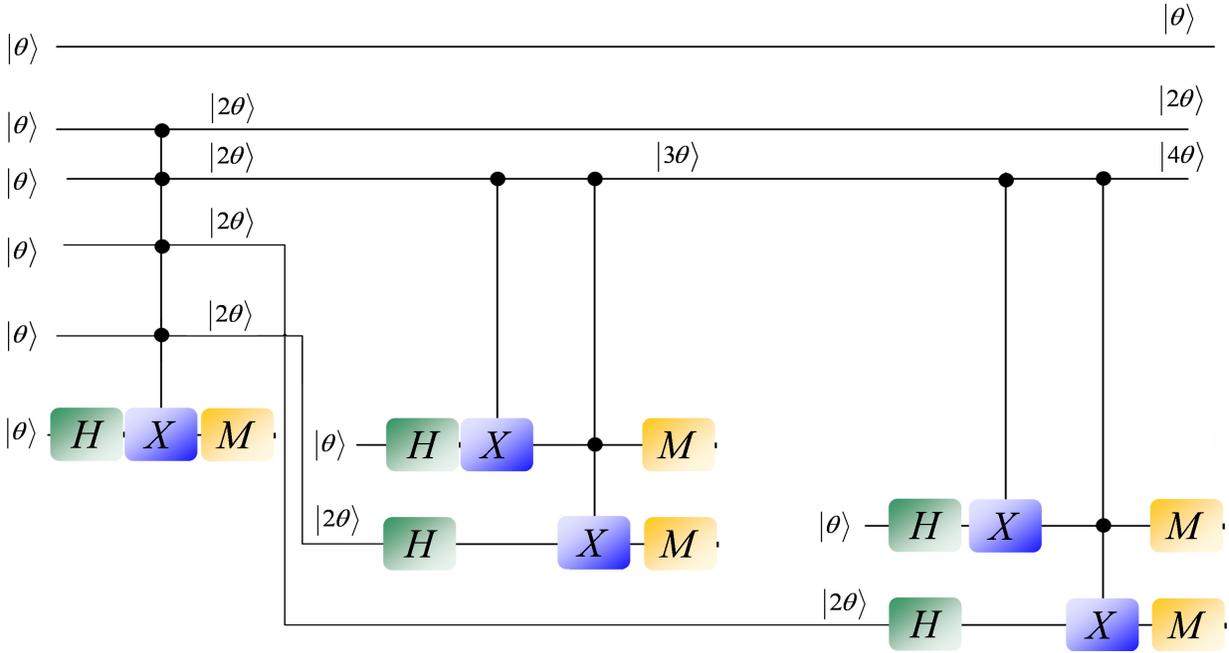

**Figure 12.** The process of generation of decoding states. The different angles are constructed from the unknown pilot states, no other information is required for Bob. After a new pilot state has been successfully generated it can be used as a target state to increase the success probability of the next rotation transformation.

For the error-correction, only the valuable $|\theta\rangle^{Pilot} = \otimes_{i=1}^{l} |2^{i-1}\theta\rangle$, i.e., the "2 power" pilot states can be used in the error correction, the generation of the other (intermediate) pilot states between these pilot states are unavoidable for the generation of these angles. (For example, $|3\theta\rangle$ will not be part of the final, $l$-length qubit string.) As follows, from the unknown "2 power" pilot states can be generated by the proposed quantum circuit, using only standard Hadamard and CNOT gates, and the $r$-length string $\otimes_{i=1}^{r} |\theta\rangle_i$ with unknown pilot channel output states. The success probability increases in every step, the final success probability of generating $|2^{l-1}\theta\rangle$ is

$$p = 1 - (1/2)^{l-2}.$$

Fig. 13 displays the generation process of $|2^{l-1}\theta\rangle$ from $|(2^{l-1}-1)\theta\rangle$, which is also an unknown state, as well as the states of pilot string from which the state will be generated.

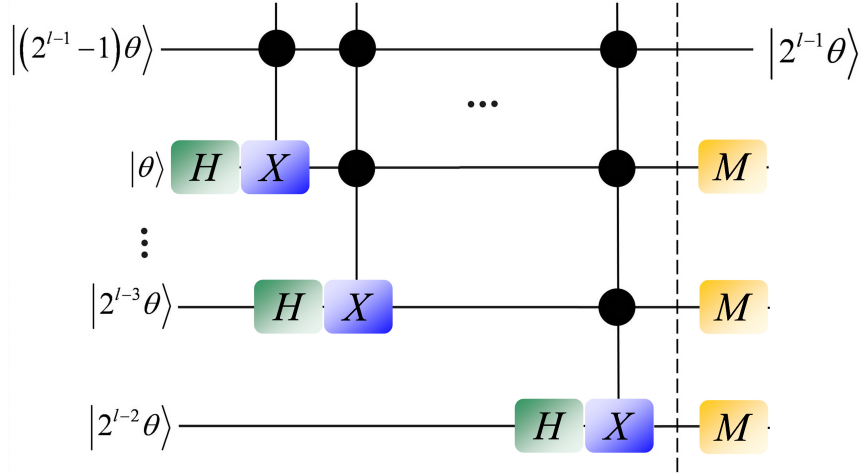

**Figure 13.** Generation of the final state of the pilot string. After this state is ready, Bob can correct an arbitrary number of data qubits with a probability $p = 1 - (1/2)^l$.

Bob rotates in every step $|\theta\rangle$ by an angle $\theta$, hence to get the final state $|2^{l-1}\theta\rangle$ from $|\theta\rangle$, he needs to apply $2^{l-1} - 1$ rotations. This process does not decrease the efficiency of our scheme, since the value of $l$ can be chosen to be several orders of magnitude lower than the number of data qubits $n$. Moreover, this error-correction scheme does not require redundant coding on the level of a single qubit, it can be carried out without the transmission of redundant qubits through the quantum channel.

These results conclude the proof of Theorem 4.

■

The result on the required number of pilot states for the generation of the valuable pilot string is summarized in Theorem 5.

## E. Required Number of Channel Output Pilot States

In the previous section we proved that it is possible to generate the set of valuable (and unknown) pilot states from the set of unknown pilot states. Here we give an exact measure to the number of channel output pilot states and the achievable success probability of the error correction.

**Theorem 5.** *The linearly increasing number of pilot states ensures the exponentially increasing success probability of the pilot error-correction.*

*Proof.* According to working mechanism of the circuit shown in Fig. 13, and the defined pilot output states in see (36) and (39), to generate the $l$-length pilot qubit string $\otimes_{i=1}^{l} |2^{i-1}\theta\rangle = |\theta\rangle \otimes \cdots \otimes |2^{l-1}\theta\rangle$, $r$ pilot states $|\theta\rangle$ needed:

$$r = 2^{l-2}(l-1) + 2^{l-3}(l-2) + \\ \cdots + 2^2 \cdot 3 + 2^1 \cdot 2 + 2^0 \cdot 1 + 1 + l - 1,$$

The required number of $|\theta\rangle$ for some type of pilot states and the success probability of the error-correction are shown in Table 1.

| $l$ | Required number of $r$ | Success probability: $p = 1 - (1/2)^l$ |
|---|---|---|
| $l=2$ | $r=3$ | $p=0.75$ |
| $l=3$ | $r=8$ | $p=0.875$ |
| $l=4$ | $r=21$ | $p=0.9375$ |
| $l=5$ | $r=54$ | $p=0.96875$ |
| $l=6$ | $r=135$ | $p=0.984375$ |

**Table 1.** Required number of generation of pilot qubits and the reachable success probability in the error correction process.

The decoding success probability in function of the length of the pilot qubit string is illustrated in Fig. 14.

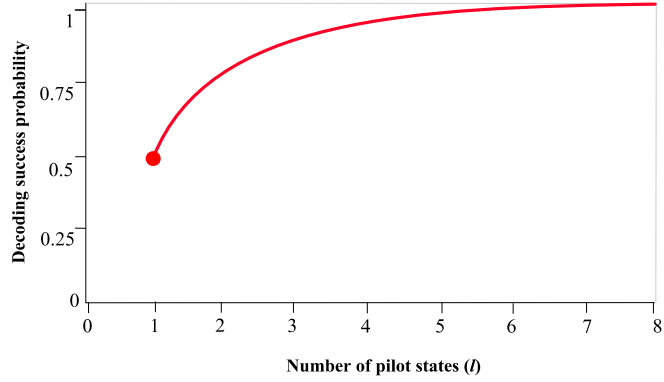

**Figure 14.** The decoding success probability increases exponentially as the number of pilot qubits increases linearly. The maximization of success probability requires only a minimal number of pilot qubits states.

The pilot error-correction process requires only a minimal number of pilot qubit states which assures the efficiency and the minimal redundancy of the technique.

But this process does not decrease the efficiency of the scheme, since the value of $l$ can be chosen to be several orders of magnitude lower than the number of data qubits $n$. After receiving $54$ pilot qubits $|\theta\rangle$ from Alice, Bob can construct a $5$-length pilot qubit string to correct arbitrary number of data qubits with success probability $0.96875$. To summarize, the linearly increasing numbers of $r$ and $l$ ensures the exponentially increasing $p = 1 - \left(1/2\right)^l$ success probability of the pilot quantum error-correction.

These results conclude the proof of Theorem 5.

∎

The result on the multi qubit error-correcting is summarized in Theorem 6.

## F. The Multi Error-Correction Process

In Theorem 6 we extend the results of Theorem 3 to the correction of an arbitrary length data string $|\psi_A\rangle_{OUT}^{Data} = \otimes_{i=1}^{n}|d_i\rangle$, where $n \to \infty$.

**Theorem 6.** *The error correction quantum circuit can achieve the correction of $n$ data states $\left|\psi_A\right\rangle_{OUT}^{Data} = \otimes_{i=1}^{n}\left|d_i\right\rangle$ simultaneously, where $n \gg l$ is an arbitrarily high number, where $n \rightarrow \infty$.*

*Proof.* The number $n$ depends on the length of the occurring of the same error on the quantum channel, our error-correction scheme works in this "pilot region". We show how Bob could achieve this for an arbitrarily long input data system with $n$ unknown input data quantum states.

If Bob receives an $n$-qubit length damaged string $\left|\psi_A\right\rangle_{OUT}^{Data} = \otimes_{i=1}^{n}\left|d_i\right\rangle = \left|d_1\right\rangle \otimes \ldots \otimes \left|d_n\right\rangle$, see (37), then he can correct all the data qubits with the $l$-qubit length pilot string $\otimes_{i=1}^{l}\left|2^{i-1}\theta\right\rangle$ (which string was generated from the $r$-length string $\left|\varphi\right\rangle_{OUT}^{Pilot} = \otimes_{i=1}^{r}\left|\theta\right\rangle_i$, see (36) and (39)) *simultaneously*, as depicted in Fig. 15. Using our method, all the rotation transformations are completed by an $l$-length multi-qubit pilot string, therefore every error correction transformation $U_\theta^\dagger$ on the corresponding damaged qubit $\left|d_i\right\rangle$ of the data qubit string can be implemented with a success probability $1 - \left(1/2\right)^l$. The outcome of the quantum circuit can be expressed in the following form

$$\frac{1}{\sqrt{2^l}}\left(\left(\sqrt{2^l-1}\right)U_\theta^{(2^l-1)\dagger}\left|\psi_A\right\rangle_{OUT}^{Data} \otimes \left|right\right\rangle + U_\theta\left|\psi_A\right\rangle_{OUT}^{Data} \otimes \left|wrong\right\rangle\right), \quad (64)$$

since in this case we have an $n$-length data string.

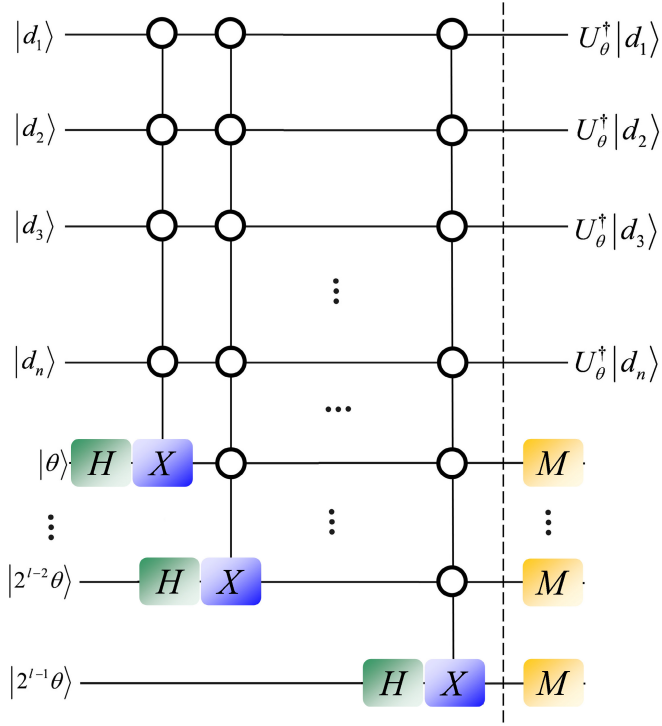

**Figure 15.** The correction of an $n$-length data string with an $l$-length multi-qubit pilot string.

The $n$-length $|\psi_A\rangle_{OUT}^{Data}$ data state can be expressed as

$$|\psi_A\rangle_{OUT}^{Data} = \sum_{k=0}^{2^n-1} c_k |k\rangle, \tag{65}$$

where $c_k$ is the complex coefficient. The output of the quantum circuit of Fig. 15 can be evaluated as follows:

$$\begin{aligned} U_\theta^\dagger |\psi_A\rangle_{OUT}^{Data} &= U_\theta^\dagger \sum_{k=0}^{2^n-1} c_k |k\rangle \\ &= \frac{1}{\sqrt{2^l}} \begin{pmatrix} \sqrt{2^l-1} U_\theta^{(2^l-1)\dagger} \sum_{k=0}^{2^n-1} c_k |k\rangle \otimes |right\rangle \\ + U_\theta \sum_{k=0}^{2^n-1} c_k |k\rangle \otimes |wrong\rangle \end{pmatrix} \\ &= \sum_{k=0}^{2^n-1} c_k \begin{pmatrix} \sqrt{2^l-1} U_\theta^{(2^l-1)\dagger} |k\rangle \otimes |right\rangle \\ + U_\theta |k\rangle \otimes |wrong\rangle \end{pmatrix}. \end{aligned} \tag{66}$$

As can be concluded, the success probability $\sqrt{2^l-1}$ is independent of the length of the $\left|\psi_A\right\rangle_{OUT}^{Data}$ data qubit string. As follows, the success probability only depends on the length $l$ of the pilot states.

These results conclude the proof of Theorem 6.

∎

## VI. Application of Pilot Quantum Error-Correction

In this section we show, that our encoding scheme can be applied in a realistic space-earth quantum communication system. The efficiency of the scheme depends on the time parameter $T$, for which the quantum channel can be assumed to be in a stationary state. The velocity at which the angle $\theta$ changes depends on the speed (and therefore on the orbit) of the satellite. The time parameter $T$ also differs for a low and high-orbit satellite systems [16]. In an experimental space-earth quantum communication system the values of $\theta$ changes fast, but as we show, the proposed pilot strategy can also be applied in these cases. (*Note*: From the viewpoint of the efficiency of pilot coding scheme, the geostationary systems would mean the optimal solution. In a high-orbit GEO (Geostationary Earth Orbit) system the rotation angle changes slow and the quantum channel takes the desired stationary state for longer time $T$, in comparison to lower orbits.) Based on these arguments, in the next subsection the performance of the pilot coding scheme will be analyzed for a realistic MEO (Medium Earth Orbit) channel implementation, where the rotation angle changes fast.

## A. Medium Earth Orbit Implementation

Assuming a Medium Earth Orbit space-earth quantum communication system with satellites above 2,000 kilometers of the Earth and between the geostationary orbit (35,786 kilometers), the time parameter during rotation $\theta$ can be taken to constant [15] is

$$T_{crit.} = 0.5 \text{ sec}. \tag{67}$$

Using a laser source with repetition rate [15]

$$f_{source} = 100 \text{ MHZ}, \tag{68}$$

during the given time-slot $T = [0, T_{crit.} = 0.5 \text{ sec}]$ the number $\Sigma$ of generated qubits is

$$\Sigma = 5 \cdot 10^7 = 50 \text{ million qubits}/T. \tag{69}$$

The attenuation parameter $\Delta$ differs in the up-link and down-link directions and also depends on the properties of the physical apparatus (such as the telescope size, motion speed, etc.).

An averaged value of the attenuation parameter for an MEO space-earth quantum channel [15] can be taken to

$$\Delta = 5 \cdot 10^{-5}. \tag{70}$$

From these parameters, the number $N = n + r$ of transmitted qubits during time-slot $T$ can be expressed as

$$\begin{aligned} N &= \Sigma \cdot \Delta = 5 \cdot 10^7 \text{ qubits}/T_{crit.} \cdot \left(5 \cdot 10^{-5}\right) \\ &= n + r = 2500 \text{ qubits}. \end{aligned} \tag{71}$$

The *D redundancy* of the code is evaluated as follows:

$$D = \frac{r}{(r+n)} = \frac{r}{N}, \tag{72}$$

where the value of $r$ is chosen according to the desired decoding success probability.

In Table 2 we summarized the number of maximal transmittable qubits using the given time-slot $T_{crit.} = 0.5$ sec and attenuation parameter $\Delta = 5 \cdot 10^{-5}$, which values can be used for an experimental space-earth quantum communication system.

Each stream characterized with the same the time-slot $T_{crit.} = 0.5$ sec and attenuation parameter $\Delta = 5 \cdot 10^{-5}$.

| Laser repetition rate: $(f_{source})$ | Generated qubits* per T: $(\Sigma)$ | Maximal transmittable qubits: $(N = n+r)$ | Corrected data qubits $(N - r = n)$ with probability p=0.96875 (l=5, r=54) | Redundancy (D) [%] |
|---|---|---|---|---|
| 100 MHZ | $5 \cdot 10^7$ | $25 \cdot 10^2$ | $N = 2446$ | 2.16% |
| 500 MHZ | $2.5 \cdot 10^8$ | $12.5 \cdot 10^3$ | $N = 12446$ | 0.43% |
| 1 GHZ | $5 \cdot 10^8$ | $25 \cdot 10^3$ | $N = 24946$ | 0.21% |
| 5 GHZ | $2.5 \cdot 10^9$ | $12.5 \cdot 10^4$ | $N = 124946$ | 0.04% |
| 10 GHZ | $5 \cdot 10^9$ | $25 \cdot 10^4$ | $N = 249946$ | 0.021% |

**Table 2.** The number of maximal transmittable and correctable data qubits during time slot $T$ with success probability $p$=0.96875 assuming MEO satellites. (*Note: The generated qubits can carry both classical or quantum information. If Alice wants to send classical information, the data block is a simple quantum repetition code, while in case of quantum information, the data block is filled with random qubits.*)

As follows from Table 2, the pilot strategy can be applied with very high efficiency in space-earth quantum communications. On the other hand, the limit on the number $\Sigma$ of transmittable qubits in time slot $T$ is depends on the applied technology, i.e., on the laser source frequency $f_{source}$ and mainly on the dead-time of the detectors. In current practical space-earth satellite quantum communication systems the limit is on the detector more than on the frequency of the laser. These practical issues can be overcome by technological innovations in the very near future [15], [18-22].

## B. Capacity and Rate Calculations

According to the results of experimental demonstrations [15], [16], for the analyzed space-earth quantum communication link the depolarizing effect can be taken to small, i.e., assuming a critical time value $T_{crit.} = 0.5$ sec for the channel, in an experimental MEO setting the $p^T$ time-dependent depolarization parameter of $\mathcal{N}_{depol.}^T$ for $T \in [0, T_{crit.}]$ can be chosen to $p^T = 0.05$ [15], [16], [35], [36] (*Note*: The value of the depolarizing parameter of the channel in the space-earth setting depends on the environment, actual weather conditions and the atmospheric parameters, i.e., it is a dynamically changing parameter. We assumed a *worst-case scenario*.). From this, the classical capacity of the $\mathcal{N}_{depol.}^T$ time-dependent depolarizing channel assuming critical time value $T_{crit.} = 0.5$ sec, can be evaluated as follows:

$$C\left(\mathcal{N}_{depol.}^T\right) = 1 - \sum_i p_i S\left(\mathcal{N}_{depol.}^T(\rho_i)\right) = 1 - H\left(\frac{1}{2} p^T\right) = 0.83134. \tag{73}$$

The quantum capacity of $\mathcal{N}_{depol.}^T$ in the MEO space-earth channel setting can be approximated by the following bound:

$$Q\left(\mathcal{N}_{depol.}^T\right) \leq 1 - 4p^T = 0.8. \tag{74}$$

For $T > T_{crit.}$ the capacities of $\mathcal{N}_{depol.}^T$ are cut down to zero, i.e.:

$$C\left(\mathcal{N}_{depol.}^T\right) = Q\left(\mathcal{N}_{depol.}^T\right) = 0. \tag{75}$$

The results for the time-dependent Medium Earth Orbit depolarizing channel are summarized in Fig. 16. The time-dependent depolarizing parameter has the constant value $p^T = 0.05$ in the time-domain $T \in [0, T_{crit.}]$. If $T > T_{crit.}$ the time-dependent depolarizing parameter takes its maximum $p^T = 1$, i.e., the classical and the quantum capacities of the channel will be equal to 0.

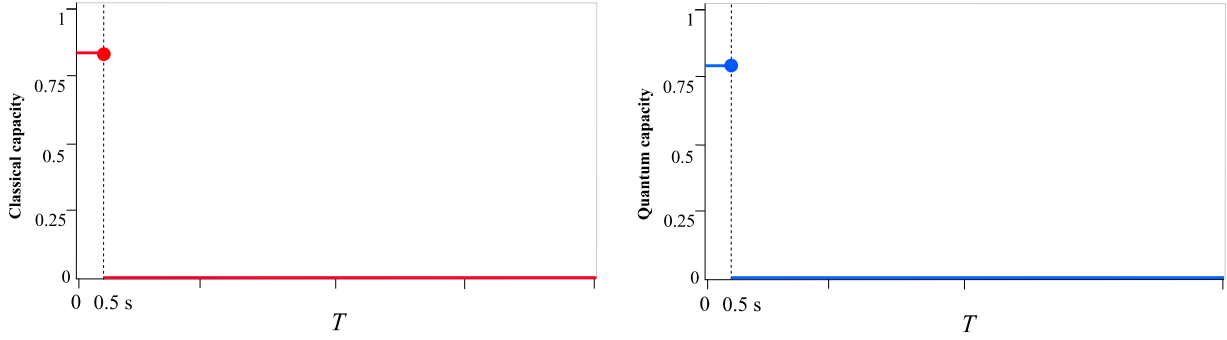

**Figure 16.** The classical and the quantum capacities of the time-dependent depolarizing channel as function of the time parameter $T$ in an experimental MEO space-earth setting.

For the rate $R$ of the pilot channel coding for an MEO system, the following conclusions can be derived. In $T = \left[0, T_{crit.} = 0.5 \text{ sec}\right]$ the rate $R$ of pilot coding scheme achieves the capacities $C\left(\mathcal{N}_{depol.}^{T}\right)$ and $Q\left(\mathcal{N}_{depol.}^{T}\right)$ of $\mathcal{N}_{depol.}^{T}$. The pilot code contains $n$ data qubits $\left|\psi_{A}\right\rangle_{IN}^{Data} = \left|\psi_{A,1}\right\rangle \otimes \ldots \otimes \left|\psi_{A,n}\right\rangle$, see (35) and $r$ pilot states $\rho_{IN}^{Pilot} = \rho_{1,IN}^{Pilot} \otimes \rho_{2,IN}^{Pilot} \otimes \ldots \otimes \rho_{r,IN}^{Pilot}$ as "redundancy", see (34), i.e., the $R_C\left(\mathcal{N}_{depol.}^{T}\right)$ classical communication rate of the code for $T = \left[0, T_{crit.}\right]$ is

$$R_C\left(\mathcal{N}_{depol.}^{T}\right) = C\left(\mathcal{N}_{depol.}^{T}\right) - \mathcal{O}\left(\frac{r}{N}\right), \qquad (76)$$

where $r$ is the number of pilot states and $N = (n+r)$ is the number of total qubits sent through over $\mathcal{N}_{depol.}^{T}$ in $T = \left[0, T_{crit.}\right]$.

The $R_Q\left(\mathcal{N}_{depol.}^{T}\right)$ quantum communication rate of the pilot code is

$$R_Q\left(\mathcal{N}_{depol.}^{T}\right) = Q\left(\mathcal{N}_{depol.}^{T}\right) - \mathcal{O}\left(\frac{r}{N}\right). \qquad (77)$$

According to the pilot code construction strategy (see Sections 4.c and 5), for sufficiently large $n$ the following relation holds

$$D = \lim_{T \to T_{crit.}} \frac{r}{N} = 0, \qquad (78)$$

which condition assures that the proposed pilot code is capacity-achieving code. The code achieves the capacity of the channel as $T \to T_{crit.}$, since after the $r$ redundant qubits are transmitted, only the value of $n$ increases, i.e., in the asymptotic setting $n \to \infty$. Due to the time-dependency of the code, in $T = [0, T_{crit.}]$ the previous result in (78) can be rephrased as

$$D = \lim_{n \to \infty} \frac{r}{(r+n)} = 0. \tag{79}$$

As follows, the rates $R$ of the code for the analyzed $\mathcal{N}_{depol.}^T$ depends on the level of redundancy (see Table 2.) Assuming $f_{source} = 100 \text{ MHZ}$, $\Sigma = 5 \cdot 10^7 = 50$ million qubits/$T$, $\Delta = 5 \cdot 10^{-5}$, $N = 2500$ qubits, $r = 54$ pilot qubits with $D = 0.0216$, the achievable rates are calculated as follows:

$$R_C\left(\mathcal{N}_{depol.}^T\right) = 0.83134 - 0.0216 = 0.80974, \tag{80}$$

and

$$R_Q\left(\mathcal{N}_{depol.}^T\right) = 0.8 - 0.0216 = 0.7784. \tag{81}$$

The code rates assuming $p_{depol.} = 0.05$ and $r = 54$ are shown in Fig. 17. The capacity-achieving property also can be observed. The rates of the code converges with exponential speed to the capacity of the time-dependent depolarizing quantum channel, which makes possible capacity-achieving classical and quantum communication over the space-earth links.

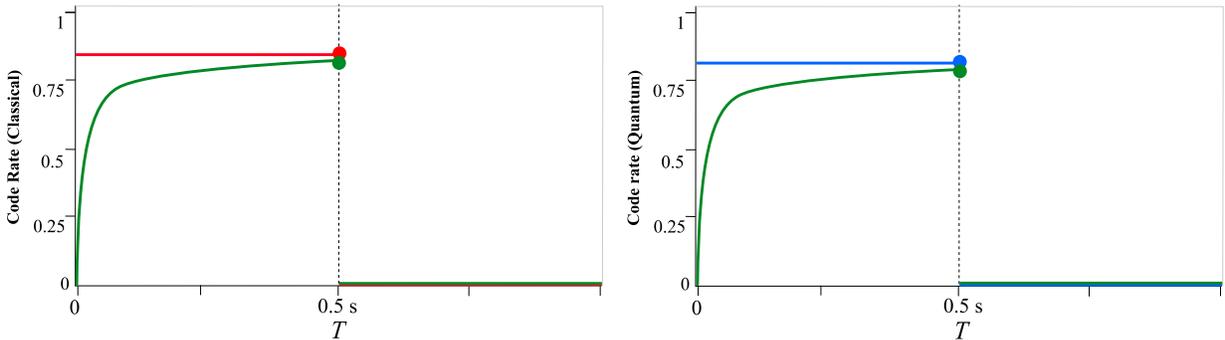

**Figure 17.** The code rates of the pilot code. The redundancy of the code decreases exponentially as the number of number of transmitted qubits increases linearly. The code makes possible capacity-achieving classical and quantum communication over the time-dependent slightly depolarizing channel. The speed of converge is equal for the classical and quantum case, since the level of redundancy of the code depends only on the number of transmitted qubits.

Using the proposed $\mathcal{N}_{depol.}^{T}$ time-dependent depolarization channel, high classical and quantum communication rates can be obtained in the time domain $T = \left[0, T_{crit.}\right]$ for an MEO setting. The $\mathcal{N}_{depol.}^{T}$ time-dependent channel in the MEO setting has very small critical depolarization parameter $p_{crit.}$, in the critical time-domain $T = \left[0, T_{crit.}\right]$. In the pilot coding scheme we do not use the elements of standard error-correction techniques (stabilizer codes). The error-correction is achieved by the pilot states, in comparison to the techniques of repetition coding, random coding, or other "well-known" standard quantum-error correction methods. As we have demonstrated, the rate of the code is determined only by the redundancy of the code. The redundancy of the code decreases exponentially with the increasing number of transmitted qubits. Using the redundant qubits (the so-called pilot states) a complete $n$ length blockcode can be corrected, with the code rates given above in (80) and (81). According to our knowledge, no similar code construction scheme was demonstrated in the literature before our results, thus the complete proof of the achievable rates of the code over arbitrary quantum channels is still open.

# VII. Conclusions

In this paper we have introduced a new quantum-error correction scheme. The protocol uses quantum states to capture and store the unknown error of the quantum channel. The error correction requires only a very small number of quantum states sent out with the data qubits - called pilot states. The pilot states characterize the unknown rotation that occurs in the polarization of the qubits. The proposed pilot quantum error-correction protocol uses a probabilistic process for the error correction, however with the generation of the required pilot states, the success probability can be increased arbitrarily high. Our solution requires the most simple quantum circuits to conduct the correction in practice, providing an easily implementable, lightweight on-the-fly error correction framework for polarization compensation.

In future work we would like to extend our method to correct any arbitrary errors in qudit systems. An important question is the determination of the time parameters of the physical quantum channels. The possibility of the correction of any non-unitary errors still keeps so many exciting results. We believe the proposed scheme can be applied to solve these issues by rather simple modifications and improvements in the error-correction circuits.

## Acknowledgements

LGY would like to thank Kamil Bradler and Markus Grassl for useful discussions and comments. The results discussed above are supported by the grant TAMOP-4.2.1/B-09/1/KMR-2010-0002, 4.2.2.B-10/1--2010-0009 and COST Action MP1006.

## References

[1] C . Bonato, M. Aspelmeyer, T. Jennewein, C. Pernechele, P. Villoresi, A. Zeilinger, Influence of satellite motion on polarization qubits in a space-earth quantum communication link. Optics Express, 14 (21):10050 – 10059, 2006.

[2] L. Hanzo, H. Haas, S. Imre, D. O'Brien, M. Rupp, L. Gyongyosi: Wireless Myths, Realities, and Futures: From 3G/4G to Optical and Quantum Wireless, *Proceedings of the IEEE*, Volume: 100, Issue: Special Centennial Issue, pp. 1853-1888.

[3] S. Imre, L. Gyongyosi: Advanced Quantum Communications – An Engineering Approach, Publisher: Wiley-IEEE Press. (In Press, 2012).

[4] S. D. Barrett, T. M. Stace, "Fault tolerant quantum computation with very high threshold for loss errors" Physical Review Letters, DOI:10.1103/PhysRevLett.105.200502 . http://arxiv.org/abs/1005.2456. (2010).

[5] C. Bennett, D. P. DiVincenzo, J. A. Smolin, and W. K. Wootters, "Mixed state entanglement and quantum error correction". Phys. Rev. A, 54(5):3824– 3851, (1996).

[6] P. Villoresi, F. Tamburini, M. Aspelmeyer, T. Jennewein, R. Ursin, C. Pernechele, G. Bianco, A. Zeilinger, C. Barbieri, Space-to-ground quantum-communication using an optical ground station: a feasibility study, SPIE proceedings Quantum Communications and Quantum Imaging II conference in Denver, July 2004.

[7] W. Dür and H.J. Briegel. Entanglement purification and quantum error correction. Rep. Prog. Phys, 70:1381–1424, (2007).

[8] D. Gottesman. Stabilizer Codes and Quantum Error Correction. PhD thesis, California Institute of Technology (arXiv:quant-ph/9705052), (1997).

[9] D. Gottesman, An Introduction to Quantum Error Correction, Quantum Computation: A Grand Mathematical Challenge for the Twenty-First Century and the Millennium, ed. S. J. Lomonaco, Jr., pp. 221-235, (2002).

[10] J. Kerckhoff, et al. "Designing Quantum Memories with Embedded Control: Photonic Circuits for Autonomous Quantum Error Correction." Physical Review Letters 105, 040502 (2010).

[11] B. Schumacher, C. M. Caves, M. A. Nielsen, and H. Barnum, "Information theoretic approach to quantum error correction and reversible measurement", Proceedings of the Royal Society of London A 454, 277-304 (1998).

[12] A. Acín, E. Jané and G. Vidal, "Optimal estimation of quantum dynamics", quant-ph/0012015. (2000).

[13] J. Kim, Y. Cheong, J. Lee, and S. Lee, Storing unitary operators in quantum states, arXiv:quant-ph/0109097v2 (2001)

[14] C. Bonato, A. Tomaello, V. D. Deppo, G. Naletto, P. Villoresi, Study of the Quantum Channel between Earth and Space for Satellite Quantum Communications

[15] P. Villoresi et al.: Experimental verification of the feasibility of a quantum channel between space and Earth. New Journal of Physics 10, 033038 (2008)

[16] C. Bonato, A. Tomaello, V. Da Deppo, G. Naletto, P. Villoresi, Some aspects on the feasibility of satellite quantum key distribution, arXiv:0903.2160v2 [quant-ph] 12 May 2009.

[17] L. Gyongyosi, S. Imre, On-the-Fly Quantum Error-Correction for Space-Earth Quantum Communication Channels, First NASA Quantum Future Technologies Conference (QFT 2012), Jan. 2012, NASA Ames Research Center, Moffett Field, California, USA.

[18] L. Gyongyosi, S. Imre, Pilot Quantum Error-Correction for Noisy Quantum Channels, Second International Conference on Quantum Error Correction (QEC11), Dec. 2011, University of Southern California, Los Angeles, USA.

[19] K. Bradler, P. Hayden, D. Touchette, and M. M. Wilde, Trade-off capacities of the quantum Hadamard channels, Journal of Mathematical Physics 51, 072201, arXiv:1001.1732v2, (2010).

[20] F. Brandao, J. Oppenheim and S. Strelchuk, "When does noise increase the quantum capacity?", arXiv:1107.4385v1 [quant-ph] (2011)

[21] L. Gyongyosi, S. Imre: Quantum Polar Coding for Noisy Optical Quantum Channels, APS DAMOP 2012 Meeting, The 43rd Annual Meeting of the APS Division of Atomic, Molecular, and Optical Physics, (American Physical Society), Jun. 2012, Anaheim, California, USA.

[22] L. Gyongyosi, S. Imre: Classical Communication with Stimulated Emission over Zero-Capacity Optical Quantum Channels, APS DAMOP 2012 Meeting, The 43rd Annual Meeting of the APS Division of Atomic, Molecular, and Optical Physics, (American Physical Society), Jun. 2012, Anaheim, California, USA.

[23] V. Buzek, M. Hillery, Universal optimal cloning of qubits and quantum registers, http://arxiv.org/abs/quant-ph/9801009

[24] A. Holevo, "The capacity of the quantum channel with general signal states", IEEE Trans. Info. Theory 44, 269 - 273 (1998).

[25] B. Schumacher and M. Westmoreland, "Sending classical information via noisy quantum channels," Phys. Rev. A, vol. 56, no. 1, pp. 131–138, (1997).

[26] S. Lloyd, "Capacity of the noisy quantum channel," Phys. Rev. A, vol. 55, pp. 1613–1622, (1997)

[27] P. Shor, "The quantum channel capacity and coherent information." lecture notes, MSRI Workshop on Quantum Computation, Available online at http://www.msri.org/publications/ln/msri/2002/quantumcrypto/shor/1/. (2002).

[28] I. Devetak, "The private classical capacity and quantum capacity of a quantum channel," IEEE Trans. Inf. Theory, vol. 51, pp. 44–55, quant-ph/0304127, (2005).

[29] B. Schumacher and M. Westmoreland, "Relative Entropy in Quantum Information Theory" 2000, LANL ArXiV e-print quant-ph/0004045, to appear in Quantum Computation and Quantum Information: A Millenium Volume , S. Lomonaco, editor (American Mathematical Society Contemporary Mathematics series), (2000).

[30] C. King, „The capacity of the quantum depolarizing channel", IEEE Trans. Info. Theory 49, no.1, 221 - 229 (2003).

[31] G. Amosov, "The strong superadditivity conjecture holds for the quantum depolarizing channel in any dimension", Phys. Rev. A 75, 2104 - 2117 (2007).

[32] D. Bruss, L. Faoro, C. Macchiavello and M. Palma, "Quantum entanglement and classical communication through a depolarizing channel", J. Mod. Opt. 47 325 (2000).

[33] N. Datta, A. S. Holevo, and Y. Suhov. Additivity for transpose depolarizing channels, http://arxiv.org/abs/quant-ph/0412034, (2004).

[34] B. Schumacher and M. Westmoreland, "Optimal Signal Ensembles", LANL ArXiV e-print quant-ph/9912122, (1999).

[35] X. Ma et al. Quantum teleportation using active feed-forward between two Canary Islands, arXiv:1205.3909v1 [quant-ph].

[36] M. Gregory, F. Heine, H. Kämpfner, R. Lange, M. Lutzer and R. Meyer, "Commercial optical inter-satellite communication at high data rates", Opt. Eng. 51, 031202 (Mar 13, 2012); http://dx.doi.org/10.1117/1.OE.51.3.031202

[37] J. Watrous, Lecture Notes in Quantum Computation, University of Calgary, (2006).